\documentclass[superscriptaddress,longbibliography,aps,prb,preprintnumbers,floatfix]{revtex4-2}

\usepackage[urlcolor=blue,colorlinks=true,citecolor=blue,linkcolor=blue,pdfstartview={FitH},bookmarks=false]{hyperref}
\usepackage[utf8]{inputenc}
\usepackage{lineno}
\usepackage{graphicx}%,comment,float}
\usepackage{xcolor,soul}
\usepackage{dcolumn}
\usepackage{bm}
\usepackage{amsmath}
\usepackage{amssymb}
\usepackage[normalem]{ulem}
\usepackage{url}
\begin{document}
\title{Large unconventional anomalous Hall effect arising from spin chirality within~domain~walls of an antiferromagnet EuZn$_2$Sb$_2$} 
\author{Karan Singh}
%\email[e-mail: ]{k.singh@intibs.pl}
\affiliation{Institute of Low Temperature and Structure Research, Polish Academy of Sciences, Wrocław, Poland}
\author{Orest Pavlosiuk}
\affiliation{Institute of Low Temperature and Structure Research, Polish Academy of Sciences, Wrocław, Poland}
\author{Shovan Dan}
\affiliation{Institute of Low Temperature and Structure Research, Polish Academy of Sciences, Wrocław, Poland}
\author{Dariusz Kaczorowski}
\affiliation{Institute of Low Temperature and Structure Research, Polish Academy of Sciences, Wrocław, Poland}
\author{Piotr Wiśniewski}
\email[e-mail: ]{p.wisniewski@intibs.pl}
\affiliation{Institute of Low Temperature and Structure Research, Polish Academy of Sciences, Wrocław, Poland}
%
%\date{\today}
\begin{abstract}
Unconventional anomalous Hall effect was observed in antiferromagnetic state of EuZn$_2$Sb$_2$. Scaling of unconventional Hall conductivity with the longitudinal conductivity, and the magnitude of Hall angle indicate spin chirality despite collinear magnetic structure. Anomalies in magnetoresistance culminate in the same fields, in which the unconventional anomalous Hall resistance has maxima. Monotonous decrease of their magnitude with increasing temperature belittles here the role of spin-fluctuations, important in isostructural compounds. These observations point to a prominent role of scalar spin chirality within domain walls, when magnetic field tilts the Eu moments. Simple calculation of such spin chirality shows it strongest in fields characteristic for anomalous magnetotransport. 
\end{abstract}
\maketitle
%
%\linenumbers
\section*{Introduction} In several Eu-based antiferromagnets from the 1:2:2 family of pnictides, e.g., EuCd$_2$As$_2$, EuCd$_2$Sb$_2$, and EuZn$_2$As$_2$, unconventional contributions to anomalous Hall effect (AHE) have recently been reported \cite{Su2020, Xu2021, Cao2022, Wang2022, Yi2023}. The origin of those contributions is still not understood well enough. Interestingly, EuCd$_2$Sb$_2$ and EuCd$_2$As$_2$ are Weyl semimetals \cite{Su2020, Soh2018, Soh2019}, while EuZn$_2$As$_2$ has the Fermi level within a narrow ($\approx 0.1\,{\rm eV}$) gap and a flat band above it \cite{Wang2022}. Spin-orbit coupling (SOC) decreases along series: EuCd$_2$Sb$_2$, EuCd$_2$As$_2$, EuZn$_2$Sb$_2$, EuZn$_2$As$_2$, whereas the appearance of topological band crossings depends on the relative strength of SOC \cite{Su2020}. It seems, however, that the difference in electronic structures induced by SOC does not significantly affect the unconventional part of anomalous Hall effect in this series, as both boundary materials, EuCd$_2$Sb$_2$ and EuZn$_2$As$_2$, exhibit pronounced unconventional AHE \cite{Su2020, Wang2022}. 
The common features of the four above-mentioned materials are: CaAl$_2$Si$_2$-type crystal structure ($P\Bar{3}m1$ space group), fairly close Néel temperatures, $T_{\rm N}$ (= 9.5, 7.4, 13.3, and 19.6\,K, respectively), and most likely the same A-type antiferromagnetic (AFM) structure, with magnetic moments of Eu$^{2+}$ ions ferromagnetically aligned within the $ab$-planes, which are stacked antiferromagnetically along ${c}$-axis \cite{Wang2022, Rahn2018}. The application of a magnetic field, \textbf{H}, leads to the canting of spins and their gradual alignment with field direction, turning the collinear magnetic structure into a non-collinear one, and above certain, sufficiently strong field, $H_{\rm sat}$, into the fully spin-polarized state \cite{Soh2018, Rahn2018, Weber2006, Blawat2022}. 
In contrast to isostructural 1:2:2 Eu-bearing compounds, the magnetoresistance and Hall effect data for EuZn$_2$Sb$_2$ have not been hitherto reported. The unconventional AHE is widespread in this family, however its origin remained unclear because collinearity of common antiferromagnetic structure seemed to exclude scalar or vector spin chirality. This gap motivated our work on high-quality single crystals of EuZn$_2$Sb$_2$, aiming to determine the origin of above mentioned anomalies.

In magnetized materials, the Hall resistivity comprises an AHE contribution, proportional to the magnetization ($M$) \cite{Karplus1954, Nagaosa2010}. Two mechanisms can be responsible for AHE: extrinsic one, related to the scattering on impurities, and the intrinsic, related to the momentum-space Berry curvature \cite{Haldane2004, Onoda2006}. 
Enhanced Berry curvature induces strong AHE and can be generated in Bloch electronic bands, when inversion symmetry or time-reversal symmetry is broken, and SOC is considerably strong \cite{Haldane2004, Liu2018}. 
The AHE induced by intrinsic mechanism, related to the Berry curvature, has been proposed for several materials, including e.g., HgCr$_2$Se$_4$, Fe$_5$Sn$_2$, Co$_3$Sn$_2$S$_2$ \cite{Liu2018, Li2020, Yang2019}. 
Significant real-space Berry curvature can occur in systems with non-coplanar spin textures, in which the so-called topological Hall effect, THE, arises due to finite scalar spin chirality defined for any three neighboring spins, ${\mathbf S}_i\; (i=1,2,3$), as  $\chi_{ijk} = {\mathbf S}_i\cdot({\mathbf S}_j\times {\mathbf S}_k)$ \cite{Nagaosa2010, Ishizuka2018, Uchida2021}. Such THE has been observed in helimagnets e.g., MnSi and MnGe \cite{Neubauer2009, Kanazawa2011}.

In antiferromagnets with collinear magnetic moments, there is neither net magnetization nor spin chirality, therefore both, AHE and THE are zero. On the other hand, several antiferromagnets with non-collinear spin structures, e.g., Mn$_5$Si$_3$ and Mn$_3$Ir, demonstrate large AHE 
\cite{Suergers2014, Chen2014}. 
In such antiferromagnets, an additional contribution to the ordinary Hall effect may alternatively be attributed to finite vector spin chirality, $\Vec{\chi}_{ij} = ({\mathbf S}_i\times {\mathbf S}_j)$ \cite{Kipp2021}, and also associated with the AFM domain walls \cite{Kim2018}. 

Our key observations for EuZn$_2$Sb$_2$ are: the large unconventional anomalous Hall effect, pronounced anomaly of magnetoresistance, and complex angular dependence of magnetoresistance. 
We propose that in AFM-ordered state all these effects stem from a common mechanism, due predominantly to the scalar spin chirality occurring within domain walls in applied magnetic field.
\section*{Sample preparation, characterization and methods of measurements}
EuZn$_2$Sb$_2$ single crystals were grown from Sb-flux by placing the starting composition, Eu:Zn:Sb, with a molar ratio of 1:2:45 in an alumina crucible. The crucible was sealed into an evacuated quartz ampule, heated up at a rate of $50^\circ$C/h to $1000^\circ$C, kept at this temperature for 20\,h and then cooled to $800^\circ$C at a rate of $2^\circ$C/h. Excess flux was removed by centrifugation. Energy dispersive X-ray spectra (EDS) taken on a few pieces of the synthesized crystals showed the stoichiometry in a good agreement with the nominal composition (see supplementary Fig.~S1a in Supplemental Material \cite{SM}). The backscattering Laue diffraction (using Laue-COS, Proto Manufacturing) provided information on the crystal orientation and quality (see supplementary Fig.~S1b \cite{SM}). It also confirmed that EuZn$_2$Sb$_2$ crystallizes in the trigonal CaAl$_2$Si$_2$-type structure, with space-group $P\bar{3}m1$, as reported previously \cite{Weber2006}.

Magnetization was measured on an oriented single crystal using a SQUID magnetometer (MPMS-XL, Quantum Design). Electrical transport measurements were performed using the Physical Property Measurements System (PPMS-14 Quantum Design) with a conventional 4-probe method. $50\,\mu{\rm m}$-thick silver wires were attached to a cuboid sample (cut from the oriented single crystals with a wire saw) with silver paint. An electric current of 1\,mA was applied along the basal \textit{ab}-plane. The effect of misalignment of the contacts was eliminated by the symmetrization and anti-symmetrization of longitudinal resistance and Hall resistance, respectively, recorded for opposite directions of applied magnetic field.

The characterization and basic physical properties of our samples (AFM with $T_{\rm N}=13.3\,$K) presented in Supplemental Material \cite{SM} (Figs.~S2 and S3, for magnetic properties and heat capacity, respectively) are in perfect agreement with an earlier report \cite{Weber2006}. 
Temperature dependence of the electrical resistivity, $\rho_{xx}$, that we measured in zero magnetic field (cf. supplementary Fig.~S4 \cite{SM}) is very similar to metallic-like one reported by May et al. \cite{May2012}. 
\section*{Hall effect and magnetoresistance} 
We measured Hall resistivity ($\rho_{xy}$) at various temperatures from 2--20\,K range, with $\textbf{j}\perp\textbf{H}\parallel c$ (Fig.~\ref{Fig1}a). 
The $M(H)$ isotherms for lowest temperatures (see Fig.~\ref{Fig1}b) show two regions of magnetic field: $H>H_{\rm sat}$, where magnetic moments are fully polarized, and $H<H_{\rm sat}$ range, in which AFM component of the magnetic structure (staggered moment) is preserved. 
The $\rho_{xy}(H>H_{\rm sat})$ is linear, thus it can be considered as a sum of the ordinary Hall resistivity, $\rho^N_{xy}=\mu_0R_0H$, and the conventional anomalous Hall resistivity, $\rho^{CA}_{xy}=R_SM$ ($R_0$ and $R_S$ are respective Hall coefficients). 
When $H<H_{\rm sat}$, there occurs pronounced non-linear contribution to $\rho_{xy}(H)$. We ascribe it to the unconventional AHE, denote as $\rho^{\rm UA}_{xy}$, and propose that its origin is associated with the canted spin structure. Thus, $\rho_{xy}(H)$ consists of three terms \cite{Kanazawa2011}: 
$\rho_{xy}=\rho^N_{xy}+\rho^{CA}_{xy}+\rho^{\rm UA}_{xy}$.  
\subsection*{Unconventional anomalous Hall resistivity}
\begin{figure}[h]
	\centering
	\includegraphics[width=0.8\textwidth]{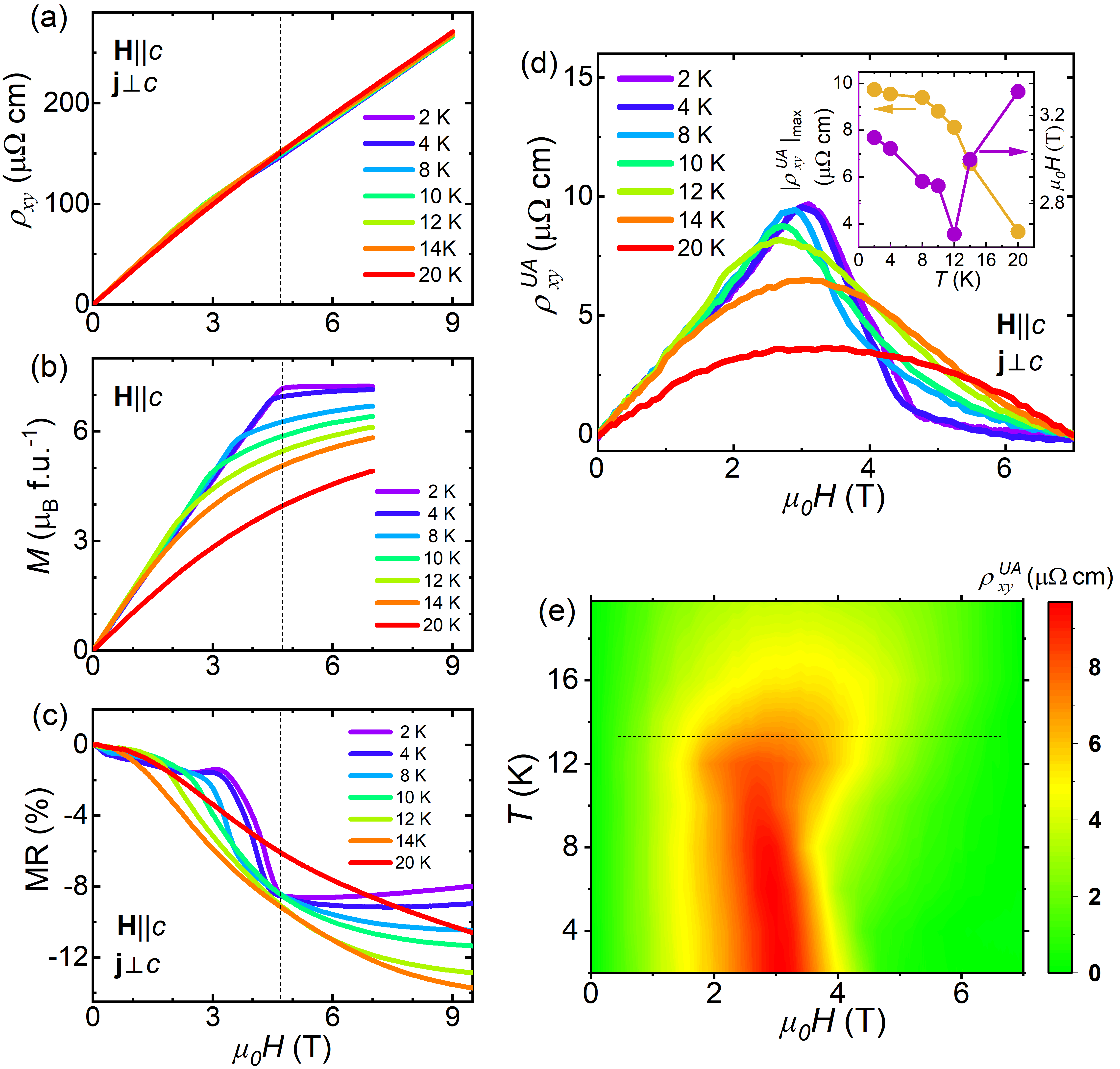}
	\caption{Magnetic field dependence of: ({\textbf a})  Hall resistivity, ({\textbf b}), magnetization and ({\textbf c}) magnetoresistance. Dashed vertical lines in ({\textbf {a-c}}) mark saturation field ($\mu_0H_{\rm sat}=4.7\,$T) at $T=2\,$K. ({\textbf d}): Unconventional anomalous Hall resistivity, $\rho^{\rm UA}_{xy}$, as a function of the applied magnetic field, at several temperatures. Inset: temperature dependence of $\rho^{\rm UA}_{xy}$ maxima (left axis) and magnetic field at which these maxima occur (right axis). ({\textbf e}): Map of $\rho^{\rm UA}_{xy}$ in the (${H,T}$) plane (dotted horizontal line marks $T_{\rm N}$).}
	\label{Fig1}
\end{figure}

We decomposed the total $\rho_{xy}(H)$ by the following steps:  
\textbf{(i)} for $T=2\,$K and 4\,K, the ordinary component of Hall resistivity $\rho^N_{xy}=R_0\mu_0H$ was estimated by linear fitting of $\rho_{xy}(H)$ in the 7--9\,T range, and slopes of these fitted lines were taken as values of $R_0$. For $T>4\,$K, $R_0$ was approximated as $\partial\rho_{xy}/\partial H$ for $\mu_0H=9$\,T. 
So estimated $R_0 \approx\,28\,\mu\Omega{\rm cm/T}$ barely changes with temperature (cf. supplementary Fig.~S5a \cite{SM}). \textbf{(ii)} This allowed us to calculate sum of anomalous components ($\rho^{CA}_{xy}+\rho^{\rm UA}_{xy})(H)$, plotted in supplementary Fig.~S5b \cite{SM} for several temperatures from the 2--20\,K range. 
At 2\,K, ($\rho^{CA}_{xy}+\rho^{\rm UA}_{xy}$)  increases with increasing $H$, attains a maximum at $\mu_0H\approx 2.8\,$T, and then decreases to become nearly constant above 4.7\,T. In the field of 7\,T all moments are aligned (cf. $M(H)$ in Fig.~\ref{Fig1}b), and ($\rho^{CA}_{xy} + \rho^{\rm UA}_{xy}$) becomes field independent. 
\textbf{(iii)} Thus, we could neglect $\rho^{\rm UA}_{xy}$ and assume that $\rho^{CA}_{xy}(7\,{\rm T}) = (\rho_{xy}-\rho^N_{xy}$). Temperature dependence of so obtained $\rho^{CA}_{xy}$ is shown in supplementary Fig.~S5c \cite{SM}. 
\textbf{(iv)} Next, we assumed that $R_S$ coefficient in $\rho^{CA}_{xy}(H)$ (usually proportional to $\rho_{xx}$ \cite{Nagaosa2010}) is field-independent at a given $T$ (based on practical independence of $\rho^{CA}_{xy}$ from the field above $H_{\rm sat}$, and overall weak field-dependence of $\rho_{xx}$). 
We used its values obtained for $\mu_0H=7\,$T as $R_s=\rho^{CA}_{xy}/M$(7\,T) to calculate the unconventional term for $H<H_{\rm sat}$, using the formula: $\rho^{\rm UA}_{xy}(H)=\rho_{xy}-\rho^{N}_{xy}(H)-R_sM(H)$. Obtained in this way $\rho^{\rm UA}_{xy}(H)$ data for several different temperatures are plotted in Fig.~\ref{Fig1}d. 

Below $T_{\rm N}$ the $\rho^{\rm UA}_{xy}(H)$ curves display distinct maxima (e.g. in $\mu_0H=2.8\,$T at $T=2\,$K). The maximum value of $9.8\,\mu\Omega{\rm cm}$ is larger than reported: $\approx0.2\,\mu\Omega{\rm cm}$, for MnGe \cite{Kanazawa2011} or $\approx2\,\mu\Omega{\rm cm}$, for CoNb$_3$S$_6$ \cite{Ghimire2018}.
The $\rho^{\rm UA}_{xy}(H)$ obtained for $T>T_{\rm N}$ are distinctly different, which suggests another underlying mechanism, most likely due to spin-fluctuations. 

For our sample, as we show in the inset to Fig.~\ref{Fig1}d, the maximum of $\rho^{\rm UA}_{xy}(H)$ isotherm ($|\rho^{\rm UA}_{xy}|_{max}$) decreases slightly with increasing temperature up to $\approx T_{\rm N}$, and rapidly drops above this temperature. 
This is in contrast to other compounds in the 1:2:2 series having the largest $|\rho^{\rm UA}_{xy}|_{max}$ at $T\approx{T_{\rm N}}$, and indicates that the most important mechanism behind $\rho^{\rm UA}_{xy}$ in AFM-state of EuZn$_2$Sb$_2$ is the magnetic texture (which is most stable at lowest $T$), and also that two different mechanisms are responsible for $\rho^{\rm UA}_{xy}$, with one of them active only at temperatures below $T_{\rm N}$. 
  
\subsection*{Comparison of unconventional anomalous Hall resistivity in isostructural compounds}
\begin{figure*}[b]
	\centering
	\includegraphics[width=0.6\textwidth]{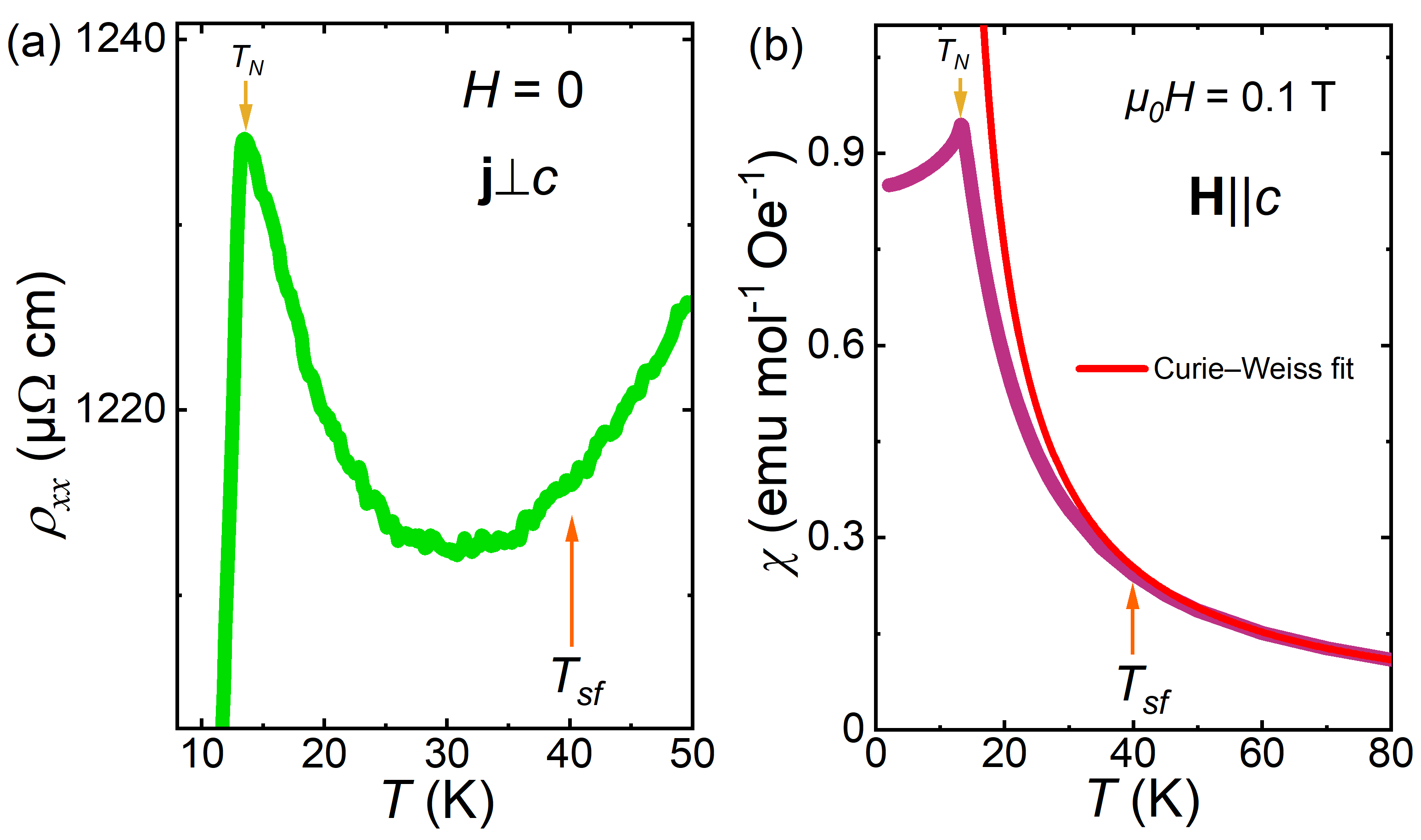}
	\caption{Temperature dependent resistivity ({\bf a}), and magnetic susceptibility ({\bf b}), of EuZn$_2$Sb$_2$ in vicinity of $T_{\rm N}$. Red line represents fitted Curie-Weiss law (see Supplemental Material \cite{SM} for details). $T_{\rm sf}$ denotes upper limit temperature for noticeable effect of spin-fluctuations.}
	\label{Fig2}
\end{figure*}

Extrema of $\rho^{\rm UA}_{xy}$ in three isostructural compounds mentioned above span vast range from -6 to $3\times10^5\,\mu\Omega{\rm cm}$, and are strongly sample-dependent \cite{Su2020, Xu2021, Cao2022, Wang2022, Yi2023,Soh2018, Soh2019}.  
In case of EuCd$_2$As$_2$ unconventional contribution has alternatively been ascribed to two different mechanisms, depending on Fermi energy (carrier concentration). Reported values of $\rho_{xy}$ (at 2\,K and 6\,T, where anomalous contributions are negligible) depend on the sample, and vary from $35\,\mu\Omega{\rm cm}$  \cite{Xu2021} to $250\,\mu\Omega{\rm cm}$ \cite{Cao2022}, corresponding to over 7-fold decrease of hole-concentration ($n_h$). In the former case (large Fermi energy) Berry curvature from Weyl points induced by fluctuating ferromagnetic domains was proposed as a driving force of $\rho^{\rm UA}_{xy}$, for the latter sample with small Fermi energy the spin canting in applied magnetic field was held responsible. 
Also, the maxima of $\rho^{\rm UA}_{xy}$ are sample dependent for EuCd$_2$As$_2$, spanning between $18\,\mu\Omega{\rm cm}$ \cite{Xu2021} and, very large $380\,\mu\Omega{\rm cm}$ \cite{Cao2022}.\\ 
Among four discussed isostructural compounds the strongest AHE is displayed by EuZn$_2$As$_2$, $\rho^{\rm UA}_{xy}\cong 3\times10^5\mu\Omega{\rm cm}$, and has been assigned to the short-range ferromagnetic order and canted magnetic moments, above and below $T_{\rm N}$, respectively, resulting in nonzero spin chirality (those results have been obtained on a sample with extremely low, but strongly $T$-dependent $n_h =3.1\times10^{15}\,{\rm cm}^{-3}$ at 2\,K and $7.5\times10^{16}\,{\rm cm}^{-3}$ at 50\,K) \cite{Yi2023}. 
Another sample of the same compound, with $T$-independent $n_h = 8.6\times10^{17}\,{\rm cm}^{-3}$, showed a large $\rho^{\rm UA}_{xy}\cong 500\, \mu\Omega{\rm cm}$ \cite{Wang2022}. 
For EuZn$_2$Sb$_2$ we found $n_h=2.2\times10^{19}\,{\rm cm}^{-3}$, intermediate between above mentioned EuCd$_2$As$_2$ samples. 
On the other hand, EuCd$_2$Sb$_2$ shows very similar $n_h=2.5\times10^{19}\,{\rm cm}^{-3}$, but its $\rho^{\rm UA}_{xy}$ is negative, having at 2\,K a local extremum of $-6\,\mu\Omega{\rm cm}$. \\
We summarized the described above literature results in Table~\ref{Tab-1} and noticed that the magnitude of $\rho^{\rm UA}_{xy}$ is correlated with both SOC and $n_h$. EuZn$_2$Sb$_2$ displays rather small $\rho^{\rm UA}_{xy}$, most likely because of rather weak effect of spin-fluctuations. In order to compare it with that in EuCd$_2$As$_2$, we estimated an upper limit temperature for noticeable effect of spin-fluctuations, $T_{\rm sf}\approx 40\,{\rm K}$, from behavior of electrical resistivity, $\rho_{xx}(T)$, and magnetic susceptibility, $\chi(T)$, as shown in Fig.~\ref{Fig2}. We also calculated relative increase of $\rho_{xx}$ due to spin-fluctuations, $\delta\rho_{xx}=(\rho_{xx}(T_{\rm N})/ \rho_{xx}(T_{\rm sf})-1)=0.016$. 
Corresponding values estimated from EuCd$_2$As$_2$ results shown in \cite{Xu2021} are much larger: $\delta\rho_{xx}>2$ and  $T_{\rm sf}\approx100\,{\rm K}$. This implicates that spin-fluctuations are much weaker, less abundant, and their effect on electrical resistivity is smaller in EuZn$_2$Sb$_2$ than in EuCd$_2$As$_2$. Nevertheless, they seem to induce the $\rho^{\rm UA}_{xy}$ we observed above $T_{\rm N}$, similar to other compounds of the series.
\begin{table}
	\caption{The magnitudes of $R_0$ (obtained from $\rho_{xy}$ in high fields, where anomalous contribution is negligible), hole concentrations $n_h$ resulting from those $R_0$ values, and extreme values of $\rho^{\rm UA}_{xy}(H)$, through the series of four 1:2:2 compounds with decreasing strength of spin orbit coupling (SOC) marked with a vertical arrow. All data, except those in the last row, were obtained at $T=2$\,K.}
	\begin{tabular}{lrrrcc}
		\hline\hline
		compound&~~$R_0\,[\Omega{\rm cm/T}]$&~~~$n_h\,[{\rm cm}^{-3}]$&$\rho^{\rm UA}_{xy}$~~&ref.~~&SOC \\
		&&&$[\mu\Omega{\rm cm}]$&&\\\hline
		\multicolumn{5}{l}{\begin{tabular}{lrrrc}
		EuCd$_2$Sb$_2$&2.5 $\times 10^{-5}$&2.5 $\times 10^{19}$&-6&\cite{Su2020} \\ \hline
		EuCd$_2$As$_2$&~~5.83 $\times 10^{-6}$&~~~1.1 $\times 10^{20}$&18&\cite{Xu2021}\\
		&7 $\times 10^{-4}$&8.9 $\times 10^{17}$&300&\cite{Soh2019}\\
		&4.17 $\times 10^{-5}$&1.5 $\times 10^{19}$&380&\cite{Cao2022}\\ \hline
		EuZn$_2$Sb$_2$&2.8 $\times 10^{-5}$&2.2 $\times 10^{19}$&10&this work\\\hline
		EuZn$_2$As$_2$&7.5 $\times 10^{-4}$&8.6 $\times 10^{17}$&500&\cite{Wang2022} \\
		&2 $\times 10^{-1}$&3.1 $\times 10^{15}$&300000&\cite{Yi2023}\\
		$T=50$\,K&8.3 $\times 10^{-3}$&3.1 $\times 10^{16}$&\textendash&\cite{Yi2023}\\		
		\end{tabular}}
		&~$\Bigg\downarrow$\\
		\hline\hline  
	\end{tabular}
	\label{Tab-1}
\end{table}

\subsection*{Scaling of unconventional Hall conductivity}  
\begin{figure}[h!]
	\centering
	\includegraphics[width=0.48\textwidth]{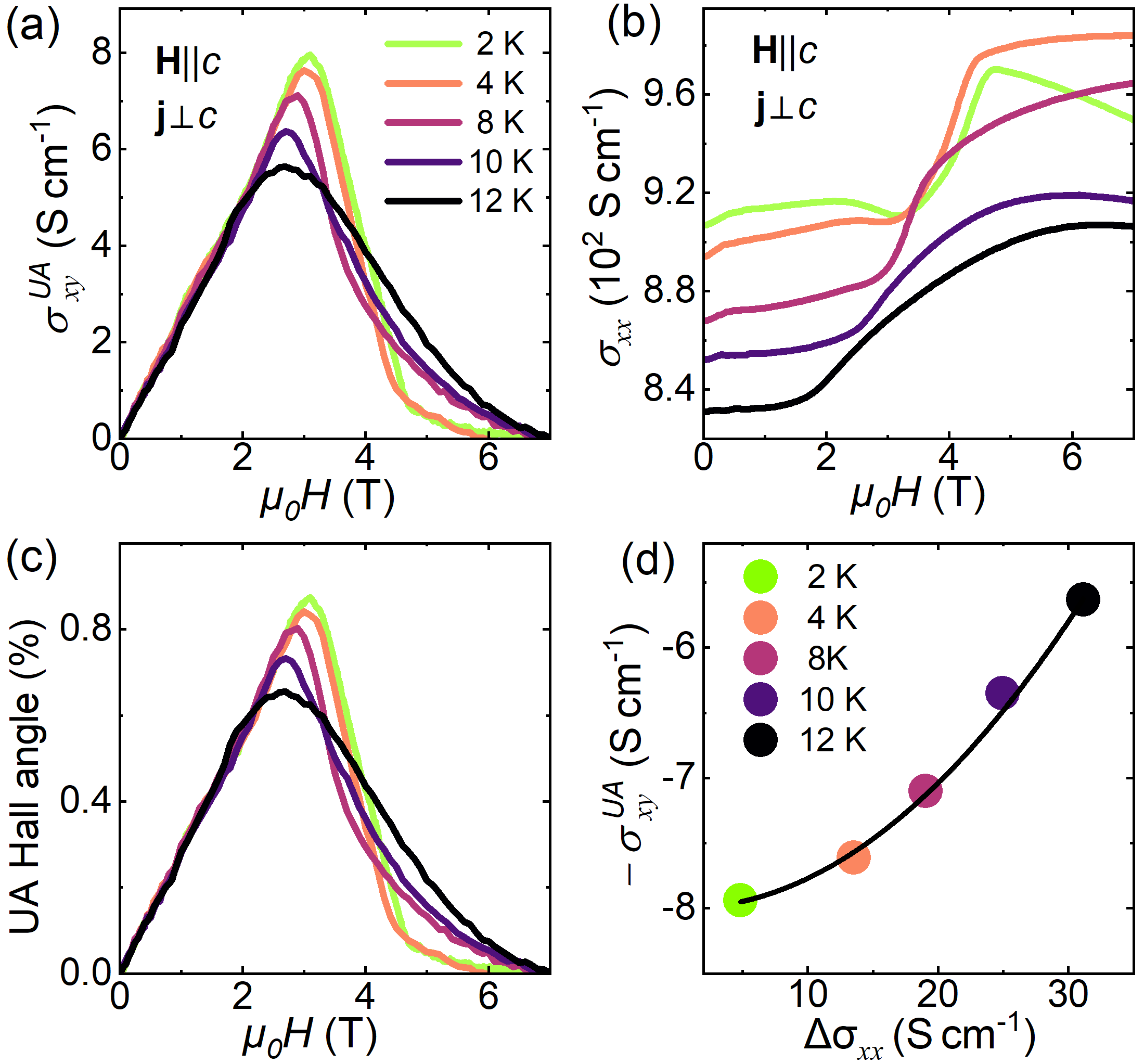}
	\caption{Magnetic field dependence of: ({\textbf a}) unconventional anomalous Hall conductivity $\sigma^{\rm UA}_{xy}$, ({\textbf b}) longitudinal conductivity $\sigma_{xx}$, and ({\textbf c}) unconventional anomalous Hall angle. ({\textbf d}): maximum value of $\sigma^{\rm UA}_{xy}$ at several temperatures, plotted as a function of $\Delta\sigma_{xx}$, and fitted with $\sigma^{\rm UA}_{xy}\propto(\Delta\sigma_{xx})^{1.8\pm0.2}$ (black solid line).}
	\label{Fig3}
\end{figure}

Next, we obtained  conductivities: 
$\sigma^{\rm UA}_{xy}={\rho^{\rm UA}_{xy}}/{(\rho^2_{xx}+\rho^2_{xy})}$ and $\sigma_{xx}={\rho_{xx}}/{(\rho^2_{xx}+\rho^2_{xy})}$, and magnetic field dependence of both components of the electrical conductivity tensor is shown in Fig.~\ref{Fig3}a and b. At 2\,K, the maximum value of anomalous Hall angle, $\theta_{AH}$ = ${\sigma^{\rm UA}_{xy}}/{\sigma_{xx}}$,
 is 0.009 (see Fig.~\ref{Fig3}c), similar to values ($\leq$0.01) estimated for Nd$_2$(Mo$_{1-x}$Nb$_x$)$_2$O$_7$ and SrFeO$_3$, in which scalar spin chirality is held responsible for AHE \cite{Ishiwata2011, Iguchi2007}. In the next step, we performed the scaling of the maximum value of anomalous Hall conductivity with the longitudinal conductivity. 
 As only one sample was studied, we used -$\sigma^{\rm UA}_{xy}$ and $\Delta\sigma_{xx}$  values obtained at several different temperatures, where $\Delta\sigma_{xx}=\sigma_{xx}(T,H)- \sigma_{xx}(T,0)$ (see Fig.~\ref{Fig3}d). Here, we subtracted the term $\sigma_{xx} (T,0)$ and used $\delta\sigma_{xx}$ for scaling, because in this way we eliminated the contribution from the magnetic disorder induced scattering, which increases, as the temperature rises. Then, we found that -$\sigma^{\rm UA}_{xy}$ is proportional to $(\Delta\sigma_{xx})^{1.8\pm0.2}$  (black solid line in Fig.~\ref{Fig3}d). This relation is close to 
 $\sigma^{AH}_{xy}\propto(\sigma_{xx})^{1.61}$ and $\sigma^{AH}_{xy}\propto(\sigma_{xx})^{1.56}$ obtained for the anomalous Hall conductivity induced by scalar spin chirality and spin-orbit interactions, respectively \cite{Iguchi2007}. 

 It has recently been proposed that also vector spin chirality can bring anomalous contribution to the Hall effect \cite{Kipp2021}, but it would cancel out in A-type AFM structure (which is common in the series of compounds we discuss). This collinear structure in zero magnetic field does not carry finite spin chirality of any kind. 
 Nevertheless, the application of magnetic field induces spin canting, which results in finite $\chi_{ijk}$, but occurring only within domain walls, as we explain below. 
\begin{figure}
\centering
\includegraphics[width=0.48\textwidth]{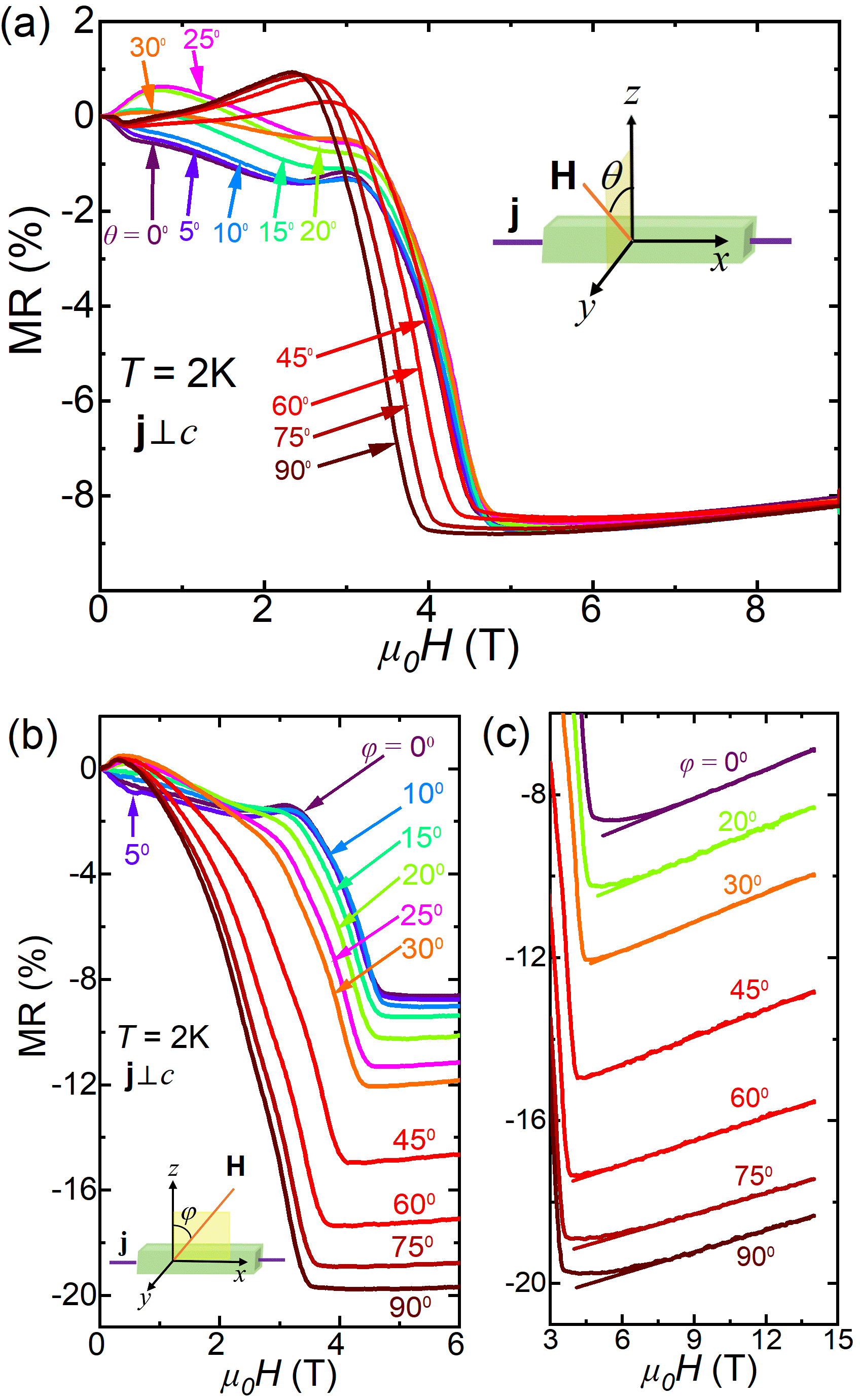}
\caption{Magnetoresistance at $T=2\,{\rm K}$ as a function of the magnetic field applied at different angles to the $c$-axis, in two geometries: with angles $\theta$ in ({\textbf a}), and $\phi$ in ({\textbf b}), between \textit{c}-axis and direction of applied field rotated in a plane $\perp{\bf j}$, and from \textit{c}-axis towards ${\bf j}$, respectively. Insets show the measurement configurations. ({\textbf c}) extended high field region of ({\textbf b}), where MR varies linearly with the field.}
\label{Fig4}
\end{figure}
\subsection*{Anomalous magnetoresistance} 
The magnetoresistance, MR $\equiv$ ($\rho_{xx}(H)$/$\rho_{xx}(0)-1)$, was measured at several temperatures, in magnetic field %(0 $< \mu_0H<10\,$T)
applied along [001] crystallographic direction ($\textbf{H}\parallel{c}$). Results are shown in Fig.~\ref{Fig1}c. Above $T_{\rm N}$, in the entire range of applied magnetic field, MR is negative and monotonically decreases with increasing magnetic field. As in many materials with large magnetic moments, such MR is mainly due to the reduction of charge carriers scattering induced by magnetic disorder. At lower temperatures, a shoulder appears in the MR($H$) curves (e.g. at $\approx3\,$T for 8\,K), with position shifting slightly towards stronger magnetic fields as $T$ decreases. 
Then, for $T\leq4\,$K, the shoulder evolves into a hump at $\approx3\,$T, which becomes more pronounced with further decreasing $T$. 
Similar anomalies in MR have been reported for EuCd$_2$As$_2$, with positions close to the magnetic fields in which the anomalous Hall effect was observed \cite{Cao2022, Yi2023, Soh2019}.\\ 
When we compared MR($H$) curves of EuZn$_2$Sb$_2$ with its $M(H)$ isotherms for $\textbf{H}\parallel{c}$ (Fig.~\ref{Fig1}b), we noticed no change in the slope of $M$ around fields for which shoulders or humps in MR($H$) were observed. 
This indicates that the origin of these anomalies is not related to any abrupt change in the magnetic structure of EuZn$_2$Sb$_2$. On the other hand, at the lowest temperatures, MR($H$) stops decreasing in the magnetic field, in which $M$ attains saturation (e.g., $\mu_0H_{\rm sat}=4.7$\,T at $T=2$\,K; cf. Fig.~\ref{Fig1}b), and a slight increase of MR is observed with further increasing of field.   
Concurrence of maxima of $\rho^{\rm UA}_{xy}(H)$ and humps of MR($H$) through whole temperature range below $T_{\rm N}$ strongly indicates that their common origin is spin texture generated by applied field. In the field range ($<\mu_0H_{\rm sat}$) where $M$ changes almost linearly, these maxima occur in fields very close to that, in which spins are canted by $\pi/4$  towards the field direction. This is also the angle of the maximal vector spin chirality resulting from two canted spins (initially antiparallel at zero field). However, vector spin chirality should cancel within a single domain, due to alternating spins in AFM structure. On the other hand, we noticed that finite $\chi_{ijk}$ occurs within domain walls and should be considered as a source of $\rho^{\rm UA}_{xy}$. 
EuZn$_2$Sb$_2$ has the largest $\mu_0H_{\rm sat}$ = 4.7\,T (at $T$ = 2\,K, if $\textbf{H}\parallel{c}$, or 3.4\,T if $\textbf{H}\perp{c}$) compared to those of EuCd$_2$Sb$_2$, EuCd$_2$As$_2$, and EuZn$_2$As$_2$: 1.6, 3.2, and 3.6\,T, respectively. This indicates the strongest AF interaction, and the largest range of field through which the canting of the spins progresses, among the series. 
The saturation field determines directly the field ($\mu_0H_{\rm sat}/\sqrt{2}$), in which modulus of the vector product of two canted spins, $|({\mathbf S}_j\times{\mathbf S}_k)|$ has a maximum. 
%This product is their vector spin chirality. 
For EuZn$_2$Sb$_2$, at 2\,K and in $\textbf{H}\parallel{c}$, $\mu_0H_{\rm sat}$/$\sqrt{2}$ is 3.3\,T, very close to 3.2\,T, where both local maximum of MR($H$) and maximum of $\rho^{\rm UA}_{xy}(H)$ occur (Fig.~\ref{Fig1}d). Also, when $\textbf{H}\perp{c}$, the MR($H$) has a maximum (cf. Fig.~\ref{Fig4}a) in such a particular field $\mu_0 H=2.4\,{\rm T}=(\mu_0H_{\rm sat}/\sqrt{2})$. 
These observations drove our attention to spin chirality as a primary source of maxima in MR and $\rho^{\rm UA}_{xy}$  of EuZn$_2$Sb$_2$. Differently than in case of AHE, there was very little discussion about MR resulting from chiral spin textures. 
Recently, however, this topic has been considered in two theoretical papers \cite{Kipp2021b,LimaF2022}. They concentrate on spin-spirals and skyrmions with spin chiralities similar to those in spin textures, that we propose below.
\vspace{-0.5cm}
\subsection*{Field-angle-dependent anomalous Hall effect}
\begin{figure*}[h]
	\centering
	\includegraphics[width=0.48\textwidth]{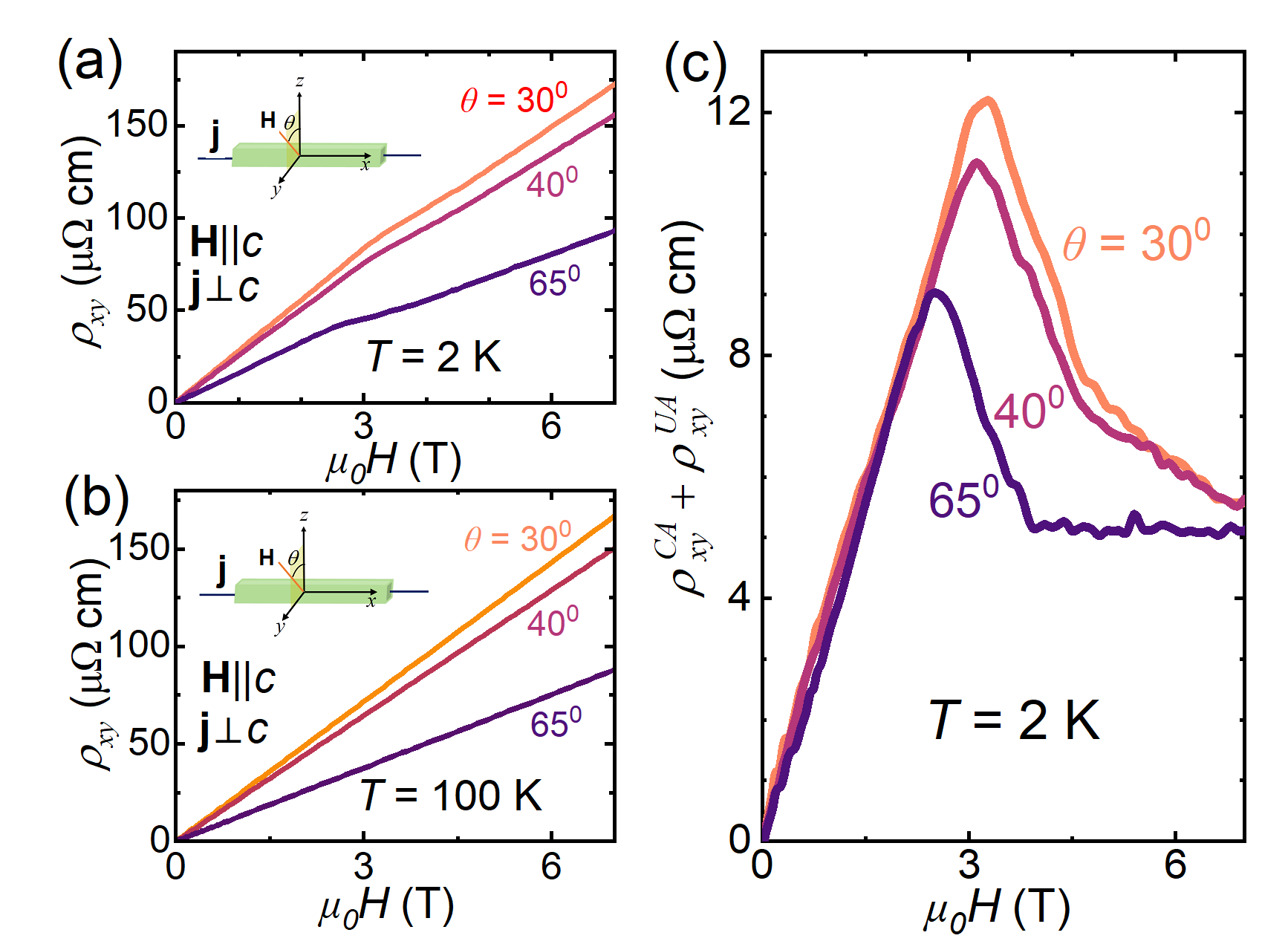}
	\caption{({\bf a}) and ({\bf b}), magnetic field dependent $\rho_{xy}(T)$ for different field angles at $T$ = 2\,K and $T$ = 100\,K, respectively. ({\bf c}) anomalous Hall contribution for different field angles, at $T=2$\,K.}
	\label{Fig5}
\end{figure*}

Figure \ref{Fig5} shows the magnetic field dependent Hall resistivity (for $\textbf{j}\perp{c}$, $\textbf{H}\perp\textbf{j}$) at different field angles from 30$^\circ$ to the 65$^\circ$, at the 2\,K and 100\,K. The anomalous Hall contribution was estimated using the relation: 
$\rho^{CA}_{xy}+\rho^{\rm UA}_{xy}=\rho_{xy}(T)- \rho^{N}_{xy}(100\;{\rm K})$. 					 
With increasing the angle, the maximum shifts from around 3\,T at 30$^\circ$ towards lower fields (see Fig.~\ref{Fig5}c). A similar shift of the hump was observed in our magnetoresistance data (see Fig.~\ref{Fig4}a). 
This corroborates the association of the hump in the magnetoresistance with the unconventional anomalous Hall effect. 
\vspace{-0.5cm}
\subsection*{Mechanism of spin chirality}  
\begin{figure}
	\centering
	\includegraphics[width=0.48\textwidth]{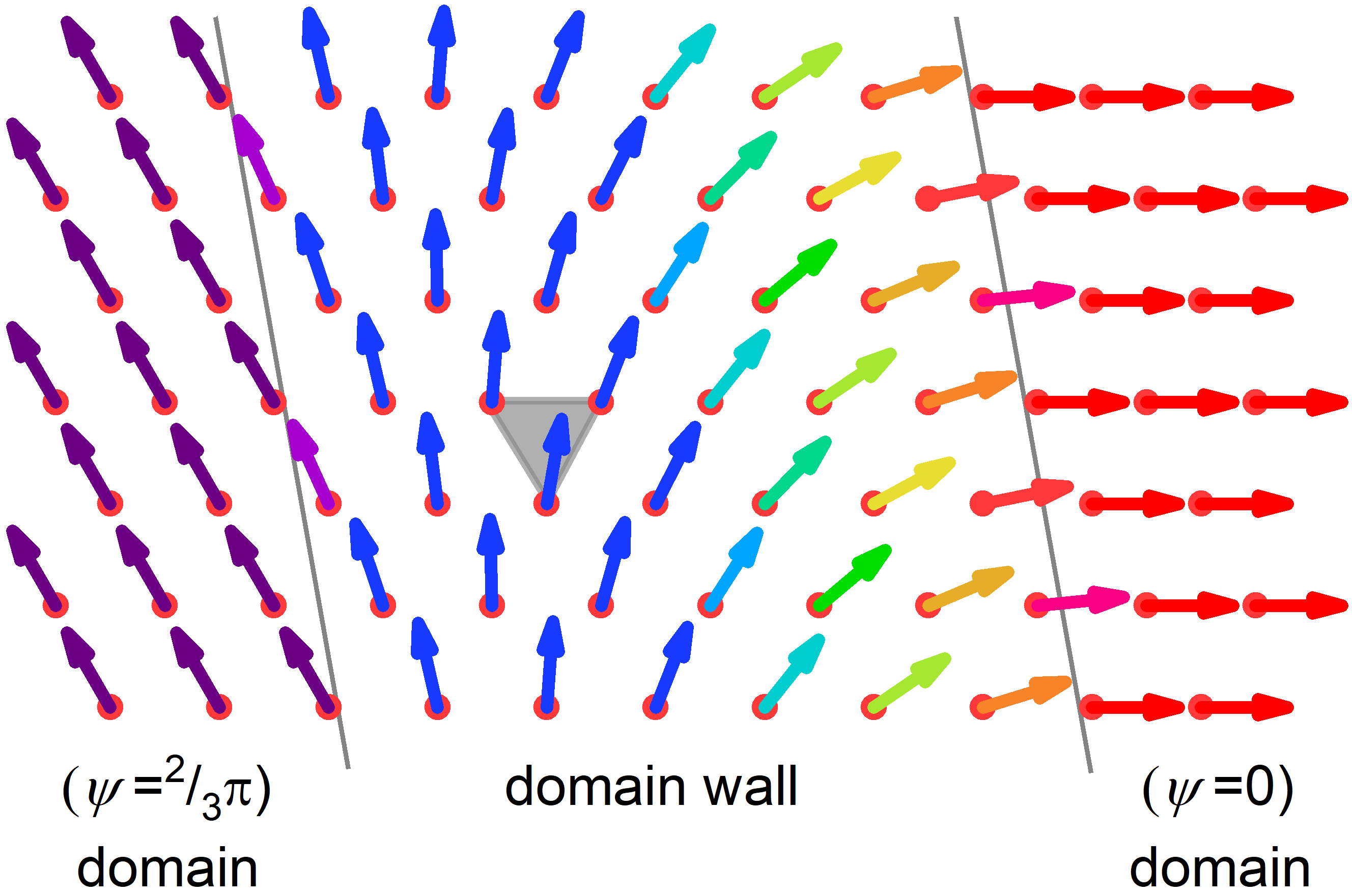}
	\caption{Spin texture within a domain wall between two domains with spin orientations differing by an azimuthal angle $\Delta\psi$ = 120$^\circ ({^2}\!/_3\pi)$. Applied magnetic field tilts all spins from \textit{ab}-plane towards \textit{c}-axis (normal to the page) by the same polar angle $\vartheta$. Every three nearest-neighbor spins (e.g. these of ions spanning the gray triangle) are non-collinear in zero-field, and become non-coplanar in applied magnetic field, which brings out finite scalar spin chirality.}
	\label{Fig6}
\end{figure}

\begin{figure*}%[h]
	\centering
	\includegraphics[width=0.49\textwidth]{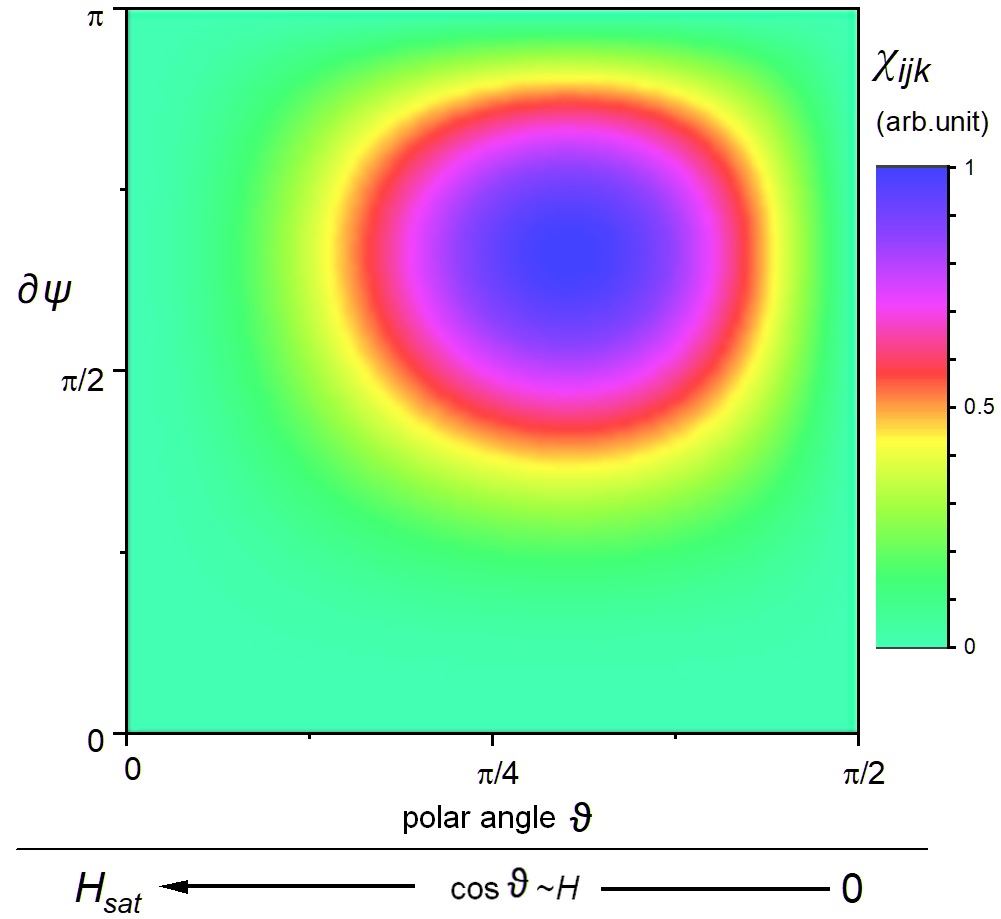}
	\caption{Map of scalar spin chirality, $\chi_{ijk}$ of three nearest-neighbor spins in the antiferromagnetic domain wall, plotted in  ($\vartheta$,$\delta\psi$)-coordinates. The polar angle of all spins is denoted with $\vartheta$, whereas $\delta\psi$ is the change of the azimuthal angle from spin to spin. The region of maximal $\chi_{ijk}$ occurs in fields $H\approx H_{\rm sat}/\sqrt{2}$.}
	\label{Fig7}
\end{figure*}

The trigonal crystal symmetry allows three equivalent antiferromagnetic domains, related by 120$^\circ$ rotation (six possible orientations of the spins). Therefore, there are three types of domain walls, between domains, which spin orientations differ by an azimuthal angle $\Delta\psi$ = 60$^\circ$, 120$^\circ$ or 180$^\circ$, respectively ($\psi$ denotes an arbitrary angle on the plane normal to $\textbf{H}$ of a projection of the spin on that plane). 
When magnetic field is applied along $c$-axis ($\textbf{H}\parallel{c}$) these domains persist. In the field applied transversely to $c$-axis ($\textbf{H}\perp{c}$), the spins quickly (in less than 0.1\,T field) undergo a flop transition and only domain walls with $\Delta\psi=180^\circ$ survive. 
In both cases, with increasing field a polar angle ($\vartheta$, between spin and $\textbf{H}$) gradually decreases, to become zero when all spins align in $H_{\rm sat}$ (sample turns into single domain). Such domain structures have been observed in EuCd$_2$As$_2$ and EuCd$_2$Sb$_2$ using resonant X-ray scattering 
\cite{Soh2018, Rahn2018}. All these domains could be considered as “quasi-ferromagnetic” domains of collinear spins, ferromagnetically coupled within trigonal layers of Eu$^{2+}$ ions. 
Of course, charge carriers moving along $ab$-plane (as in all our magnetotransport measurements) encounter domain walls, which brings additional contributions to MR and Hall resistivity. In the applied field of given strength, the spins within such domain walls have the same $\vartheta$, but azimuthal angle gradually changing by $\delta\psi$ steps across a domain wall. 
Therefore, they are non-coplanar and provide finite scalar spin chirality. We sketched an example of such spin texture in Fig.~\ref{Fig6}.
We calculated $\chi_{ijk}$ for three spins within a domain wall differing gradually by an arbitrary $\delta\psi$ (from 0 to $\pi$ range) and having the same $\vartheta$ (from 0 to $\pi$/2, which can be mapped to field $H\propto\cos\vartheta$, changing between $H_{\rm sat}$ and zero). Map of $\chi_{ijk}$ in ($\vartheta$,$\delta\psi$)-coordinates is shown in Fig.~\ref{Fig7}.
For whole range of $\delta\psi$, $\chi_{ijk}$ is finite and attains largest values for $\vartheta$ slightly above $\pi$/4, which corresponds to $H=H_{\rm sat}$/$\sqrt{2}$. 

In EuCd$_2$As$_2$ and EuZn$_2$As$_2$ maxima of $\rho^{\rm UA}_{xy}$ are huge 380, 500 or even $3\times10^5\,\mu\Omega{\rm cm}$, cf. Table~\ref{Tab-1}), but the applied fields, in which they occur are weak (0.2 and 0.15\,T, respectively), and perhaps the most significant: these maxima are strongest at 9.5\,K and 20\,K, i.e., at, or in the close vicinity of respective $T_{\rm N}$ \cite{Cao2022, Yi2023, Wang_2022}. 
In contrast: maxima of $\rho^{\rm UA}_{xy}(H)$ in EuZn$_2$Sb$_2$ reach only about $10\,\mu\Omega{\rm cm}$, but are most prominent at the lowest temperature of 2\,K, and in stronger magnetic fields ($>2/3 H_{\rm sat}$), this further implies dominating role of non-coplanar spin structure, instead of critical spin-fluctuations, or short-range order. 
%\vspace{-0.9cm}
\subsection*{Field-angle-dependent MR} 
%%%%%
\begin{figure*}
\centering
\includegraphics[width=0.48\textwidth]{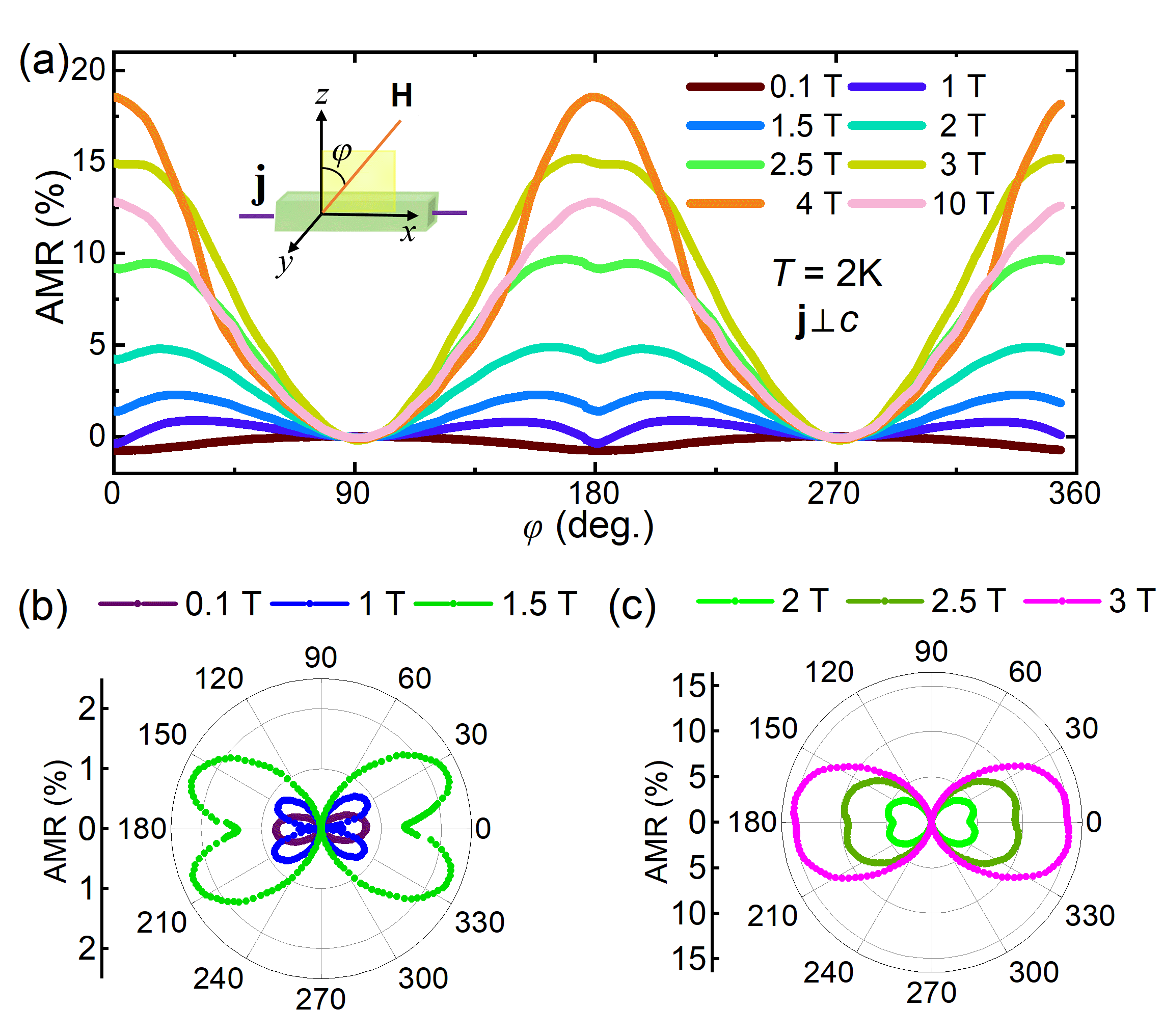}
\caption{({\bf a}) Angular dependence of the anisotropic magnetoresistance at $T$ = 2\,K for different magnetic fields. ({\bf b}) and ({\bf c}) polar plots of the data shown in panel ({\bf a}), separated in two field ranges.}
\label{Fig8}
\end{figure*}
To shed more light on the effect of the applied field on the anomaly in MR, we measured field-angle-dependent magnetoresistance in two geometries. 
In the first (see Fig.~\ref{Fig4}a), the magnetic field direction was always perpendicular to the current, $\textbf{j}$. In the second geometry, the magnetic field direction was gradually changed from transverse ($\textbf{H}\parallel{c}\perp\textbf{j}$, $\phi$ = 0$^\circ$) to longitudinal ($\textbf{H}\parallel\textbf{j}$, $\phi$ = 90$^\circ$) to the current (see Fig.~\ref{Fig4}b).  
In both cases, MR($H$) curves exhibit complex, angle-dependent behavior. The hump in MR($H$) at $\approx 3$\,T (for $\textbf{H}\parallel{c}$, and $T=2$\,K), which we associated with the unconventional AHE, in the first geometry it almost does not change its position up to $\theta=45^\circ$ (Fig.~\ref{Fig4}a), and at larger $\theta$ it shifts to slightly lower magnetic fields ($\approx2.5$\,T for $\theta=90^\circ$). This is in good agreement with the changes of anomalous Hall resistivity with the change in $\theta$ (see Fig.~\ref{Fig5}c), and simply reflects the decrease of $\mu_0H_{\rm sat}$ to 3.7\,T (and therefore of $\mu_0H_{\rm sat}/\sqrt{2}$ to 2.6\,T). 
In fields above $H_{\rm sat}$ the curves for all $\theta$ collapse on a single one, representing rather typical, weak transverse MR $\propto$ $H^2$.
In the second geometry (Fig.~\ref{Fig4}b), when $\phi$ increases, the position of the hump of MR($H$) shifts to lower magnetic fields, but for $\phi\geq$ 15$^\circ$ the hump gradually disappears and turns into less prominent shoulder. 
The strong change of MR($H$) curves with $\phi$ is due to gradual decreasing of $H_{\rm sat}$ (shifting feature related to the spin texture to lower fields) and typical anisotropic magnetoresistance, AMR$=(\rho(\phi)/\rho_\parallel)-1$, where $\rho_\parallel$=$\rho$($\phi$=90$^\circ$), behaving as $\cos2\phi$ (clearly visible for $\mu_0H$ = 10\,T, see Fig.~\ref{Fig8}a). 
Evidently, tilting of the direction of applied magnetic field from transverse to longitudinal to the current is detrimental to the mechanism inducing the hump of MR($H$) at $\approx$ $H_{\rm sat}$ /$\sqrt{2}$ (c.f. Fig.~\ref{Fig4}b). Assuming the scalar spin chirality as an important source of Berry curvature \cite{Nagaosa2010}, we note that cyclotron movement in transverse field makes the trajectories of electrons in momentum space almost closed and as a consequence their wavefunctions acquire the geometrical phase factor much more efficiently than electrons in longitudinal field, moving with almost straight trajectories \cite{Berry1984, Xiao2010}. 
Therefore, the decay of the anomaly of MR($H$) at $\approx$ $H_{\rm sat}$/$\sqrt{2}$ observed when the field changes from transverse to longitudinal is another strong argument in favor of the scalar spin chirality as principal cause of that anomaly. In fields above $H_{\rm sat}$ MR varies almost linearly with the $H$ up to the 14\,T for all $\phi$ angles as shown in Fig.~\ref{Fig4}c. The semi-classical MR varies as $\propto$ $H^2$ in low fields and then at certain fields saturates to the constant value \cite{Hu2008}. Linear MR has been reported many times and most often was attributed to the inhomogeneous mobility caused by disorder \cite{Parish2003}.

Figure \ref{Fig8}a shows anisotropic magnetoresistance (AMR$\,=[\rho_{xx}(\phi)/\rho_{xx}(90^{\circ})]-1$) recorded at $T$ = 2\,K and in different magnetic fields. For $H>H_{\rm sat}$, AMR($\phi$) behaves typically, as $\cos^2\phi$. 
For $\mu_0H$ = 0.1\,T, AMR exhibits behavior close to $\sin^2\phi$. In the intermediate range of magnetic fields, the behavior of AMR($\phi$) is more complex, with noticeable local minima at $\phi=0^\circ$ and $180^\circ$. 
The polar plot of AMR($\phi$) has a butterfly-like shape (see Fig.~\ref{Fig8}b and \ref{Fig8}c). Similar features in the AMR($\phi$) have been reported for several materials, including ZrSiS \cite{Ali2016, Voerman2019, Novak2019}, HoPtBi \cite{Pavlosiuk2020} and BaMn$_2$Bi$_2$ \cite{Huynh2019}. 
In ZrSiS it has been ascribed to near-perfect electron-hole compensation \cite{Voerman2019, Novak2019}. For EuZn$_2$Sb$_2$ that mechanism is excluded because holes are dominant carriers. However, it is noteworthy that the local minima of the AMR appear in the same field range where spin chirality induces unconventional anomalous Hall effect. 

\section*{Conclusions} We demonstrated the origin of unconventional anomalous Hall effect and pronounced anomaly in the magnetoresistance in the EuZn$_2$Sb$_2$, a collinear antiferromagnet with FM arrangement of Eu$^{2+}$ moments in the basal planes, stacked antiferromagnetically along the ${c}$-axis. 
This compound exhibits trigonal crystal symmetry, allowing three equivalent antiferromagnetic domains with moments having the azimuthal angle of 60$^\circ$, 120$^\circ$, or 180$^\circ$, respectively. In magnetic fields, spins become canted, inducing scalar spin chirality within the domain walls, having maxima for $H=H_{\rm sat}$/$\sqrt{2}$, which corresponds to a polar angle $\vartheta$ slightly above $\pi/4$. 
In the same field range, the large unconventional anomalous Hall and strong anomaly in the MR arise. Furthermore, our field-angle-dependent magnetotransport findings corroborate the conceptualization of spin chirality in this compound. \\\\
\textbf{We acknowledge} Dr.\,T. Romanova for growth of the single crystals. This work was supported by National Science Centre (NCN, Poland) under Project no. 2021/41/B/ST3/01141 (K.S., D.K., O.P. and P.W.).
%
%*Corresponding author: Piotr Wiśniewski 
%\email[e-mail:{p.wisniewski@intibs.pl}
%
%%\bibliography{biblio.bib}

\begin{thebibliography}{42}
\makeatletter
\providecommand \@ifxundefined [1]{%
	\@ifx{#1\undefined}
}%
\providecommand \@ifnum [1]{%
	\ifnum #1\expandafter \@firstoftwo
	\else \expandafter \@secondoftwo
	\fi
}%
\providecommand \@ifx [1]{%
	\ifx #1\expandafter \@firstoftwo
	\else \expandafter \@secondoftwo
	\fi
}%
\providecommand \natexlab [1]{#1}%
\providecommand \enquote  [1]{``#1''}%
\providecommand \bibnamefont  [1]{#1}%
\providecommand \bibfnamefont [1]{#1}%
\providecommand \citenamefont [1]{#1}%
\providecommand \href@noop [0]{\@secondoftwo}%
\providecommand \href [0]{\begingroup \@sanitize@url \@href}%
\providecommand \@href[1]{\@@startlink{#1}\@@href}%
\providecommand \@@href[1]{\endgroup#1\@@endlink}%
\providecommand \@sanitize@url [0]{\catcode `\\12\catcode `\$12\catcode
	`\&12\catcode `\#12\catcode `\^12\catcode `\_12\catcode `\%12\relax}%
\providecommand \@@startlink[1]{}%
\providecommand \@@endlink[0]{}%
\providecommand \url  [0]{\begingroup\@sanitize@url \@url }%
\providecommand \@url [1]{\endgroup\@href {#1}{\urlprefix }}%
\providecommand \urlprefix  [0]{URL }%
\providecommand \Eprint [0]{\href }%
\providecommand \doibase [0]{https://doi.org/}%
\providecommand \selectlanguage [0]{\@gobble}%
\providecommand \bibinfo  [0]{\@secondoftwo}%
\providecommand \bibfield  [0]{\@secondoftwo}%
\providecommand \translation [1]{[#1]}%
\providecommand \BibitemOpen [0]{}%
\providecommand \bibitemStop [0]{}%
\providecommand \bibitemNoStop [0]{.\EOS\space}%
\providecommand \EOS [0]{\spacefactor3000\relax}%
\providecommand \BibitemShut  [1]{\csname bibitem#1\endcsname}%
\let\auto@bib@innerbib\@empty
%</preamble>
\bibitem [{\citenamefont {{H. Su et al.}}(2020)}]{Su2020}%
\BibitemOpen
\bibfield  {author} {\bibinfo {author} {\bibnamefont {{H. Su et al.}}},\
}\bibfield  {title} {\bibinfo {title} {{Magnetic Exchange Induced Weyl State
			in a Semimetal EuCd$_{2}$Sb$_{2}$}},\ }\href
{https://doi.org/10.1063/1.5129467} {\bibfield  {journal} {\bibinfo
		{journal} {APL Mater.}\ }\textbf {\bibinfo {volume} {8}},\ \bibinfo {pages}
	{011109} (\bibinfo {year} {2020})}\BibitemShut {NoStop}%
\bibitem [{\citenamefont {{Y. Xu et al.}}(2021)}]{Xu2021}%
\BibitemOpen
\bibfield  {author} {\bibinfo {author} {\bibnamefont {{Y. Xu et al.}}},\
}\bibfield  {title} {\bibinfo {title} {{Unconventional Transverse Transport
			above and below the Magnetic Transition Temperature in Weyl Semimetal
			EuCd$_2$As$_2$}},\ }\href {https://doi.org/10.1103/PhysRevLett.126.076602}
{\bibfield  {journal} {\bibinfo  {journal} {Phys. Rev. Lett.}\ }\textbf
	{\bibinfo {volume} {126}},\ \bibinfo {pages} {076602} (\bibinfo {year}
	{2021})}\BibitemShut {NoStop}%
\bibitem [{\citenamefont {{X. Cao et al.}}(2022)}]{Cao2022}%
\BibitemOpen
\bibfield  {author} {\bibinfo {author} {\bibnamefont {{X. Cao et al.}}},\
}\bibfield  {title} {\bibinfo {title} {{Giant Nonlinear Anomalous Hall Effect
			Induced by Spin-Dependent Band Structure Evolution}},\ }\href
{https://doi.org/10.1103/PhysRevResearch.4.023100} {\bibfield  {journal}
	{\bibinfo  {journal} {Phys. Rev. Res.}\ }\textbf {\bibinfo {volume} {4}},\
	\bibinfo {pages} {023100} (\bibinfo {year} {2022})}\BibitemShut {NoStop}%
\bibitem [{\citenamefont {{Z.-C. Wang et al.}}(2022)}]{Wang2022}%
\BibitemOpen
\bibfield  {author} {\bibinfo {author} {\bibnamefont {{Z.-C. Wang et al.}}},\
}\bibfield  {title} {\bibinfo {title} {{Anisotropy of the Magnetic and
			Transport Properties of EuZn$_{2}$As$_{2}$}},\ }\href
{https://doi.org/10.1103/PhysRevB.105.165122} {\bibfield  {journal} {\bibinfo
		{journal} {Phys. Rev. B}\ }\textbf {\bibinfo {volume} {105}},\ \bibinfo
	{pages} {165122} (\bibinfo {year} {2022})}\BibitemShut {NoStop}%
\bibitem [{\citenamefont {{E. Yi et al.}}(2023)}]{Yi2023}%
\BibitemOpen
\bibfield  {author} {\bibinfo {author} {\bibnamefont {{E. Yi et al.}}},\
}\bibfield  {title} {\bibinfo {title} {{Topological Hall Effect Driven by
			Short-Range Magnetic Order in EuZn$_2$As$_2$}},\ }\href
{https://doi.org/10.1103/PhysRevB.107.035142} {\bibfield  {journal} {\bibinfo
		{journal} {Phys. Rev. B}\ }\textbf {\bibinfo {volume} {107}},\ \bibinfo
	{pages} {035142} (\bibinfo {year} {2023})}\BibitemShut {NoStop}%
\bibitem [{\citenamefont {Soh}\ \emph {et~al.}(2018)\citenamefont {Soh},
	\citenamefont {Donnerer}, \citenamefont {Hughes}, \citenamefont {Schierle},
	\citenamefont {Weschke}, \citenamefont {Prabhakaran},\ and\ \citenamefont
	{Boothroyd}}]{Soh2018}%
\BibitemOpen
\bibfield  {author} {\bibinfo {author} {\bibfnamefont {J.-R.}\ \bibnamefont
		{Soh}}, \bibinfo {author} {\bibfnamefont {C.}~\bibnamefont {Donnerer}},
	\bibinfo {author} {\bibfnamefont {K.~M.}\ \bibnamefont {Hughes}}, \bibinfo
	{author} {\bibfnamefont {E.}~\bibnamefont {Schierle}}, \bibinfo {author}
	{\bibfnamefont {E.}~\bibnamefont {Weschke}}, \bibinfo {author} {\bibfnamefont
		{D.}~\bibnamefont {Prabhakaran}},\ and\ \bibinfo {author} {\bibfnamefont
		{A.~T.}\ \bibnamefont {Boothroyd}},\ }\bibfield  {title} {\bibinfo {title}
	{{Magnetic and Electronic Structure of the Layered Rare-Earth Pnictide
			EuCd$_2$Sb$_2$}},\ }\href {https://doi.org/10.1103/PhysRevB.98.064419}
{\bibfield  {journal} {\bibinfo  {journal} {Phys. Rev. B}\ }\textbf {\bibinfo
		{volume} {98}},\ \bibinfo {pages} {064419} (\bibinfo {year}
	{2018})}\BibitemShut {NoStop}%
\bibitem [{\citenamefont {{J.-R. Soh et al.}}(2019)}]{Soh2019}%
\BibitemOpen
\bibfield  {author} {\bibinfo {author} {\bibnamefont {{J.-R. Soh et al.}}},\
}\bibfield  {title} {\bibinfo {title} {{Ideal Weyl Semimetal Induced by
			Magnetic Exchange}},\ }\href {https://doi.org/10.1103/PhysRevB.100.201102}
{\bibfield  {journal} {\bibinfo  {journal} {Phys. Rev. B}\ }\textbf {\bibinfo
		{volume} {100}},\ \bibinfo {pages} {201102(R)} (\bibinfo {year}
	{2019})}\BibitemShut {NoStop}%
\bibitem [{\citenamefont {Rahn}\ \emph {et~al.}(2018)\citenamefont {Rahn},
	\citenamefont {Soh}, \citenamefont {Francoual}, \citenamefont {Veiga},
	\citenamefont {Strempfer}, \citenamefont {Mardegan}, \citenamefont {Yan},
	\citenamefont {Guo}, \citenamefont {Shi},\ and\ \citenamefont
	{Boothroyd}}]{Rahn2018}%
\BibitemOpen
\bibfield  {author} {\bibinfo {author} {\bibfnamefont {M.~C.}\ \bibnamefont
		{Rahn}}, \bibinfo {author} {\bibfnamefont {J.~R.}\ \bibnamefont {Soh}},
	\bibinfo {author} {\bibfnamefont {S.}~\bibnamefont {Francoual}}, \bibinfo
	{author} {\bibfnamefont {L.~S.~I.}\ \bibnamefont {Veiga}}, \bibinfo {author}
	{\bibfnamefont {J.}~\bibnamefont {Strempfer}}, \bibinfo {author}
	{\bibfnamefont {J.}~\bibnamefont {Mardegan}}, \bibinfo {author}
	{\bibfnamefont {D.~Y.}\ \bibnamefont {Yan}}, \bibinfo {author} {\bibfnamefont
		{Y.~F.}\ \bibnamefont {Guo}}, \bibinfo {author} {\bibfnamefont {Y.~G.}\
		\bibnamefont {Shi}},\ and\ \bibinfo {author} {\bibfnamefont {A.~T.}\
		\bibnamefont {Boothroyd}},\ }\bibfield  {title} {\bibinfo {title} {{Coupling
			of Magnetic Order and Charge Transport in the Candidate Dirac Semimetal
			EuCd$_2$As$_2$}},\ }\href {https://doi.org/10.1103/PhysRevB.97.214422}
{\bibfield  {journal} {\bibinfo  {journal} {Phys. Rev. B}\ }\textbf {\bibinfo
		{volume} {97}},\ \bibinfo {pages} {214422} (\bibinfo {year}
	{2018})}\BibitemShut {NoStop}%
\bibitem [{\citenamefont {Weber}\ \emph {et~al.}(2006)\citenamefont {Weber},
	\citenamefont {Cosceev}, \citenamefont {Drobnik}, \citenamefont {Faißt},
	\citenamefont {Grube}, \citenamefont {Nateprov}, \citenamefont {Pfleiderer},
	\citenamefont {Uhlarz},\ and\ \citenamefont {v.~Löhneysen}}]{Weber2006}%
\BibitemOpen
\bibfield  {author} {\bibinfo {author} {\bibfnamefont {F.}~\bibnamefont
		{Weber}}, \bibinfo {author} {\bibfnamefont {A.}~\bibnamefont {Cosceev}},
	\bibinfo {author} {\bibfnamefont {S.}~\bibnamefont {Drobnik}}, \bibinfo
	{author} {\bibfnamefont {A.}~\bibnamefont {Faißt}}, \bibinfo {author}
	{\bibfnamefont {K.}~\bibnamefont {Grube}}, \bibinfo {author} {\bibfnamefont
		{A.}~\bibnamefont {Nateprov}}, \bibinfo {author} {\bibfnamefont
		{C.}~\bibnamefont {Pfleiderer}}, \bibinfo {author} {\bibfnamefont
		{M.}~\bibnamefont {Uhlarz}},\ and\ \bibinfo {author} {\bibfnamefont
		{H.}~\bibnamefont {v.~Löhneysen}},\ }\bibfield  {title} {\bibinfo {title}
	{{Low-Temperature Properties and Magnetic Order of EuZn$_2$Sb$_2$}},\ }\href
{https://doi.org/10.1103/PhysRevB.73.014427} {\bibfield  {journal} {\bibinfo
		{journal} {Phys. Rev. B}\ }\textbf {\bibinfo {volume} {73}},\ \bibinfo
	{pages} {014427} (\bibinfo {year} {2006})}\BibitemShut {NoStop}%
\bibitem [{\citenamefont {Blawat}\ \emph {et~al.}(2022)\citenamefont {Blawat},
	\citenamefont {Marshall}, \citenamefont {Singleton}, \citenamefont {Feng},
	\citenamefont {Cao}, \citenamefont {Xie},\ and\ \citenamefont
	{Jin}}]{Blawat2022}%
\BibitemOpen
\bibfield  {author} {\bibinfo {author} {\bibfnamefont {J.}~\bibnamefont
		{Blawat}}, \bibinfo {author} {\bibfnamefont {M.}~\bibnamefont {Marshall}},
	\bibinfo {author} {\bibfnamefont {J.}~\bibnamefont {Singleton}}, \bibinfo
	{author} {\bibfnamefont {E.}~\bibnamefont {Feng}}, \bibinfo {author}
	{\bibfnamefont {H.}~\bibnamefont {Cao}}, \bibinfo {author} {\bibfnamefont
		{W.}~\bibnamefont {Xie}},\ and\ \bibinfo {author} {\bibfnamefont
		{R.}~\bibnamefont {Jin}},\ }\bibfield  {title} {\bibinfo {title} {{Unusual
			Electrical and Magnetic Properties in Layered EuZn$_2$As$_2$}},\ }\href
{https://doi.org/10.1002/qute.202200012} {\bibfield  {journal} {\bibinfo
		{journal} {Adv. Quant. Tech.}\ }\textbf {\bibinfo {volume} {5}},\ \bibinfo
	{pages} {2200012} (\bibinfo {year} {2022})}\BibitemShut {NoStop}%
\bibitem [{\citenamefont {Karplus}\ and\ \citenamefont
	{Luttinger}(1954)}]{Karplus1954}%
\BibitemOpen
\bibfield  {author} {\bibinfo {author} {\bibfnamefont {R.}~\bibnamefont
		{Karplus}}\ and\ \bibinfo {author} {\bibfnamefont {J.~M.}\ \bibnamefont
		{Luttinger}},\ }\bibfield  {title} {\bibinfo {title} {{Hall Effect in
			Ferromagnetics}},\ }\href {https://doi.org/10.1103/PhysRev.95.1154}
{\bibfield  {journal} {\bibinfo  {journal} {Phys. Rev.}\ }\textbf {\bibinfo
		{volume} {95}},\ \bibinfo {pages} {1154} (\bibinfo {year}
	{1954})}\BibitemShut {NoStop}%
\bibitem [{\citenamefont {Nagaosa}\ \emph {et~al.}(2010)\citenamefont
	{Nagaosa}, \citenamefont {Sinova}, \citenamefont {Onoda}, \citenamefont
	{MacDonald},\ and\ \citenamefont {Ong}}]{Nagaosa2010}%
\BibitemOpen
\bibfield  {author} {\bibinfo {author} {\bibfnamefont {N.}~\bibnamefont
		{Nagaosa}}, \bibinfo {author} {\bibfnamefont {J.}~\bibnamefont {Sinova}},
	\bibinfo {author} {\bibfnamefont {S.}~\bibnamefont {Onoda}}, \bibinfo
	{author} {\bibfnamefont {A.~H.}\ \bibnamefont {MacDonald}},\ and\ \bibinfo
	{author} {\bibfnamefont {N.~P.}\ \bibnamefont {Ong}},\ }\bibfield  {title}
{\bibinfo {title} {{Anomalous Hall Effect}},\ }\href
{https://doi.org/10.1103/RevModPhys.82.1539} {\bibfield  {journal} {\bibinfo
		{journal} {Rev. Mod. Phys.}\ }\textbf {\bibinfo {volume} {82}},\ \bibinfo
	{pages} {1539} (\bibinfo {year} {2010})}\BibitemShut {NoStop}%
\bibitem [{\citenamefont {Haldane}(2004)}]{Haldane2004}%
\BibitemOpen
\bibfield  {author} {\bibinfo {author} {\bibfnamefont {F.~D.~M.}\
		\bibnamefont {Haldane}},\ }\bibfield  {title} {\bibinfo {title} {{Berry
			Curvature on the Fermi Surface: Anomalous Hall Effect as a Topological
			Fermi-Liquid Property}},\ }\href
{https://doi.org/10.1103/PhysRevLett.93.206602} {\bibfield  {journal}
	{\bibinfo  {journal} {Phys. Rev. Lett.}\ }\textbf {\bibinfo {volume} {93}},\
	\bibinfo {pages} {206602} (\bibinfo {year} {2004})}\BibitemShut {NoStop}%
\bibitem [{\citenamefont {Onoda}\ \emph {et~al.}(2006)\citenamefont {Onoda},
	\citenamefont {Sugimoto},\ and\ \citenamefont {Nagaosa}}]{Onoda2006}%
\BibitemOpen
\bibfield  {author} {\bibinfo {author} {\bibfnamefont {S.}~\bibnamefont
		{Onoda}}, \bibinfo {author} {\bibfnamefont {N.}~\bibnamefont {Sugimoto}},\
	and\ \bibinfo {author} {\bibfnamefont {N.}~\bibnamefont {Nagaosa}},\
}\bibfield  {title} {\bibinfo {title} {{Intrinsic versus Extrinsic Anomalous
			Hall Effect in Ferromagnets}},\ }\href
{https://doi.org/10.1103/PhysRevLett.97.126602} {\bibfield  {journal}
	{\bibinfo  {journal} {Phys. Rev. Lett.}\ }\textbf {\bibinfo {volume} {97}},\
	\bibinfo {pages} {126602} (\bibinfo {year} {2006})}\BibitemShut {NoStop}%
\bibitem [{\citenamefont {{E. Liu et al.}}(2018)}]{Liu2018}%
\BibitemOpen
\bibfield  {author} {\bibinfo {author} {\bibnamefont {{E. Liu et al.}}},\
}\bibfield  {title} {\bibinfo {title} {{Giant Anomalous Hall Effect in a
			Ferromagnetic Kagome-Lattice Semimetal}},\ }\href
{https://doi.org/10.1038/s41567-018-0234-5} {\bibfield  {journal} {\bibinfo
		{journal} {Nat. Phys.}\ }\textbf {\bibinfo {volume} {14}},\ \bibinfo {pages}
	{1125} (\bibinfo {year} {2018})}\BibitemShut {NoStop}%
\bibitem [{\citenamefont {{H. Li et al.}}(2020)}]{Li2020}%
\BibitemOpen
\bibfield  {author} {\bibinfo {author} {\bibnamefont {{H. Li et al.}}},\
}\bibfield  {title} {\bibinfo {title} {{Large Anomalous Hall Effect in a
			Hexagonal Ferromagnetic Fe$_5$Sn$_3$}},\ }\href
{https://doi.org/10.1103/PhysRevB.101.140409} {\bibfield  {journal} {\bibinfo
		{journal} {Phys. Rev. B}\ }\textbf {\bibinfo {volume} {101}},\ \bibinfo
	{pages} {140409(R)} (\bibinfo {year} {2020})}\BibitemShut {NoStop}%
\bibitem [{\citenamefont {Yang}\ \emph {et~al.}(2019)\citenamefont {Yang},
	\citenamefont {Li}, \citenamefont {Lin}, \citenamefont {Yi}, \citenamefont
	{Shi}, \citenamefont {Culcer},\ and\ \citenamefont {Li}}]{Yang2019}%
\BibitemOpen
\bibfield  {author} {\bibinfo {author} {\bibfnamefont {S.}~\bibnamefont
		{Yang}}, \bibinfo {author} {\bibfnamefont {Z.}~\bibnamefont {Li}}, \bibinfo
	{author} {\bibfnamefont {C.}~\bibnamefont {Lin}}, \bibinfo {author}
	{\bibfnamefont {C.}~\bibnamefont {Yi}}, \bibinfo {author} {\bibfnamefont
		{Y.}~\bibnamefont {Shi}}, \bibinfo {author} {\bibfnamefont {D.}~\bibnamefont
		{Culcer}},\ and\ \bibinfo {author} {\bibfnamefont {Y.}~\bibnamefont {Li}},\
}\bibfield  {title} {\bibinfo {title} {{Unconventional Temperature Dependence
			of the Anomalous Hall Effect in HgCr$_2$Se$_4$}},\ }\href
{https://doi.org/10.1103/PhysRevLett.123.096601} {\bibfield  {journal}
	{\bibinfo  {journal} {Phys. Rev. Lett.}\ }\textbf {\bibinfo {volume} {123}},\
	\bibinfo {pages} {096601} (\bibinfo {year} {2019})}\BibitemShut {NoStop}%
\bibitem [{\citenamefont {Ishizuka}\ and\ \citenamefont
	{Nagaosa}(2018)}]{Ishizuka2018}%
\BibitemOpen
\bibfield  {author} {\bibinfo {author} {\bibfnamefont {H.}~\bibnamefont
		{Ishizuka}}\ and\ \bibinfo {author} {\bibfnamefont {N.}~\bibnamefont
		{Nagaosa}},\ }\bibfield  {title} {\bibinfo {title} {{Spin Chirality Induced
			Skew Scattering and Anomalous Hall Effect in Chiral Magnets}},\ }\href
{https://doi.org/10.1126/sciadv.aap9962} {\bibfield  {journal} {\bibinfo
		{journal} {Sci. Adv.}\ }\textbf {\bibinfo {volume} {4}},\ \bibinfo {pages}
	{eaap9962} (\bibinfo {year} {2018})}\BibitemShut {NoStop}%
\bibitem [{\citenamefont {{M. Uchida et al.}}(2021)}]{Uchida2021}%
\BibitemOpen
\bibfield  {author} {\bibinfo {author} {\bibnamefont {{M. Uchida et al.}}},\
}\bibfield  {title} {\bibinfo {title} {{Above-Ordering-Temperature Large
			Anomalous Hall Effect in a Triangular-Lattice Magnetic Semiconductor}},\
}\href {https://doi.org/10.1126/sciadv.abl5381} {\bibfield  {journal}
	{\bibinfo  {journal} {Sci. Adv.}\ }\textbf {\bibinfo {volume} {7}},\ \bibinfo
	{pages} {eabl5381} (\bibinfo {year} {2021})}\BibitemShut {NoStop}%
\bibitem [{\citenamefont {Neubauer}\ \emph {et~al.}(2009)\citenamefont
	{Neubauer}, \citenamefont {Pfleiderer}, \citenamefont {Binz}, \citenamefont
	{Rosch}, \citenamefont {Ritz}, \citenamefont {Niklowitz},\ and\ \citenamefont
	{Böni}}]{Neubauer2009}%
\BibitemOpen
\bibfield  {author} {\bibinfo {author} {\bibfnamefont {A.}~\bibnamefont
		{Neubauer}}, \bibinfo {author} {\bibfnamefont {C.}~\bibnamefont
		{Pfleiderer}}, \bibinfo {author} {\bibfnamefont {B.}~\bibnamefont {Binz}},
	\bibinfo {author} {\bibfnamefont {A.}~\bibnamefont {Rosch}}, \bibinfo
	{author} {\bibfnamefont {R.}~\bibnamefont {Ritz}}, \bibinfo {author}
	{\bibfnamefont {P.~G.}\ \bibnamefont {Niklowitz}},\ and\ \bibinfo {author}
	{\bibfnamefont {P.}~\bibnamefont {Böni}},\ }\bibfield  {title} {\bibinfo
	{title} {{Hall Effect in the A Phase of MnSi}},\ }\href
{https://doi.org/10.1103/PhysRevLett.102.186602} {\bibfield  {journal}
	{\bibinfo  {journal} {Phys. Rev. Lett.}\ }\textbf {\bibinfo {volume} {102}},\
	\bibinfo {pages} {186602} (\bibinfo {year} {2009})}\BibitemShut {NoStop}%
\bibitem [{\citenamefont {Kanazawa}\ \emph {et~al.}(2011)\citenamefont
	{Kanazawa}, \citenamefont {Onose}, \citenamefont {Arima}, \citenamefont
	{Okuyama}, \citenamefont {Ohoyama}, \citenamefont {Wakimoto}, \citenamefont
	{Kakurai}, \citenamefont {Ishiwata},\ and\ \citenamefont
	{Tokura}}]{Kanazawa2011}%
\BibitemOpen
\bibfield  {author} {\bibinfo {author} {\bibfnamefont {N.}~\bibnamefont
		{Kanazawa}}, \bibinfo {author} {\bibfnamefont {Y.}~\bibnamefont {Onose}},
	\bibinfo {author} {\bibfnamefont {T.}~\bibnamefont {Arima}}, \bibinfo
	{author} {\bibfnamefont {D.}~\bibnamefont {Okuyama}}, \bibinfo {author}
	{\bibfnamefont {K.}~\bibnamefont {Ohoyama}}, \bibinfo {author} {\bibfnamefont
		{S.}~\bibnamefont {Wakimoto}}, \bibinfo {author} {\bibfnamefont
		{K.}~\bibnamefont {Kakurai}}, \bibinfo {author} {\bibfnamefont
		{S.}~\bibnamefont {Ishiwata}},\ and\ \bibinfo {author} {\bibfnamefont
		{Y.}~\bibnamefont {Tokura}},\ }\bibfield  {title} {\bibinfo {title} {{Large
			Topological Hall Effect in a Short-Period Helimagnet MnGe}},\ }\href
{https://doi.org/10.1103/PhysRevLett.106.156603} {\bibfield  {journal}
	{\bibinfo  {journal} {Phys. Rev. Lett.}\ }\textbf {\bibinfo {volume} {106}},\
	\bibinfo {pages} {156603} (\bibinfo {year} {2011})}\BibitemShut {NoStop}%
\bibitem [{\citenamefont {Sürgers}\ \emph {et~al.}(2014)\citenamefont
	{Sürgers}, \citenamefont {Fischer}, \citenamefont {Winkel},\ and\
	\citenamefont {Löhneysen}}]{Suergers2014}%
\BibitemOpen
\bibfield  {author} {\bibinfo {author} {\bibfnamefont {C.}~\bibnamefont
		{Sürgers}}, \bibinfo {author} {\bibfnamefont {G.}~\bibnamefont {Fischer}},
	\bibinfo {author} {\bibfnamefont {P.}~\bibnamefont {Winkel}},\ and\ \bibinfo
	{author} {\bibfnamefont {H.~V.}\ \bibnamefont {Löhneysen}},\ }\bibfield
{title} {\bibinfo {title} {{Large Topological Hall Effect in the
			Non-Collinear Phase of an Antiferromagnet}},\ }\href
{https://doi.org/10.1038/ncomms4400} {\bibfield  {journal} {\bibinfo
		{journal} {Nat. Commun.}\ }\textbf {\bibinfo {volume} {5}},\ \bibinfo {pages}
	{3400} (\bibinfo {year} {2014})}\BibitemShut {NoStop}%
\bibitem [{\citenamefont {Chen}\ \emph {et~al.}(2014)\citenamefont {Chen},
	\citenamefont {Niu},\ and\ \citenamefont {MacDonald}}]{Chen2014}%
\BibitemOpen
\bibfield  {author} {\bibinfo {author} {\bibfnamefont {H.}~\bibnamefont
		{Chen}}, \bibinfo {author} {\bibfnamefont {Q.}~\bibnamefont {Niu}},\ and\
	\bibinfo {author} {\bibfnamefont {A.~H.}\ \bibnamefont {MacDonald}},\
}\bibfield  {title} {\bibinfo {title} {{Anomalous Hall Effect Arising from
			Noncollinear Antiferromagnetism}},\ }\href
{https://doi.org/10.1103/PhysRevLett.112.017205} {\bibfield  {journal}
	{\bibinfo  {journal} {Phys. Rev. Lett.}\ }\textbf {\bibinfo {volume} {112}},\
	\bibinfo {pages} {017205} (\bibinfo {year} {2014})}\BibitemShut {NoStop}%
\bibitem [{\citenamefont {{J. Kipp et al.}}(2021)}]{Kipp2021}%
\BibitemOpen
\bibfield  {author} {\bibinfo {author} {\bibnamefont {{J. Kipp et al.}}},\
}\bibfield  {title} {\bibinfo {title} {{The Chiral Hall Effect in Canted
			Ferromagnets and Antiferromagnets}},\ }\href
{https://doi.org/10.1038/s42005-021-00587-3} {\bibfield  {journal} {\bibinfo
		{journal} {Commun. Phys.}\ }\textbf {\bibinfo {volume} {4}},\ \bibinfo
	{pages} {1} (\bibinfo {year} {2021})}\BibitemShut {NoStop}%
\bibitem [{\citenamefont {{W. J. Kim et al.}}(2018)}]{Kim2018}%
\BibitemOpen
\bibfield  {author} {\bibinfo {author} {\bibnamefont {{W. J. Kim et al.}}},\
}\bibfield  {title} {\bibinfo {title} {{Unconventional Anomalous Hall Effect 
		from Antiferromagnetic Domain Walls of Nd$_2$Ir$_2$O$_7$ thin films}},\
}\href {https://doi.org/10.1103/PhysRevB.98.125103} {\bibfield  {journal}
	{\bibinfo  {journal} {Phys. Rev. B}\ }\textbf {\bibinfo {volume} {98}},\
	\bibinfo {pages} {125103} (\bibinfo {year} {2018})}\BibitemShut {NoStop}%
\bibitem [{SM()}]{SM}%
\BibitemOpen
\href@noop {} {\bibinfo {title} {See Supplemental Material below for sample characterization (including EDS spectrum, Laue diffractogram, and data on magnetization, heat capacity, electrical 
	resistivity), and temperature variation of the ordinary and anomalous Hall effect.}}\BibitemShut {Stop}%
\bibitem [{\citenamefont {May}\ \emph {et~al.}(2012)\citenamefont {May},
	\citenamefont {McGuire}, \citenamefont {Ma}, \citenamefont {Delaire},
	\citenamefont {Huq},\ and\ \citenamefont {Custelcean}}]{May2012}%
\BibitemOpen
\bibfield  {author} {\bibinfo {author} {\bibfnamefont {A.~F.}\ \bibnamefont
		{May}}, \bibinfo {author} {\bibfnamefont {M.~A.}\ \bibnamefont {McGuire}},
	\bibinfo {author} {\bibfnamefont {J.}~\bibnamefont {Ma}}, \bibinfo {author}
	{\bibfnamefont {O.}~\bibnamefont {Delaire}}, \bibinfo {author} {\bibfnamefont
		{A.}~\bibnamefont {Huq}},\ and\ \bibinfo {author} {\bibfnamefont
		{R.}~\bibnamefont {Custelcean}},\ }\bibfield  {title} {\bibinfo {title}
	{{Properties of Single Crystalline \textit{A}Zn$_2$Sb$_2$ (\textit{A} =
			Ca,Eu,Yb)}},\ }\href {https://doi.org/10.1063/1.3681817} {\bibfield
	{journal} {\bibinfo  {journal} {J. Appl. Phys.}\ }\textbf {\bibinfo {volume}
		{111}},\ \bibinfo {pages} {033708} (\bibinfo {year} {2012})}\BibitemShut
{NoStop}%
\bibitem [{\citenamefont {Ghimire}\ \emph {et~al.}(2018)\citenamefont
	{Ghimire}, \citenamefont {Botana}, \citenamefont {Jiang}, \citenamefont
	{Zhang}, \citenamefont {Chen},\ and\ \citenamefont {Mitchell}}]{Ghimire2018}%
\BibitemOpen
\bibfield  {author} {\bibinfo {author} {\bibfnamefont {N.~J.}\ \bibnamefont
		{Ghimire}}, \bibinfo {author} {\bibfnamefont {A.~S.}\ \bibnamefont {Botana}},
	\bibinfo {author} {\bibfnamefont {J.~S.}\ \bibnamefont {Jiang}}, \bibinfo
	{author} {\bibfnamefont {J.}~\bibnamefont {Zhang}}, \bibinfo {author}
	{\bibfnamefont {Y.-S.}\ \bibnamefont {Chen}},\ and\ \bibinfo {author}
	{\bibfnamefont {J.~F.}\ \bibnamefont {Mitchell}},\ }\bibfield  {title}
{\bibinfo {title} {{Large Anomalous Hall Effect in the Chiral-Lattice
			Antiferromagnet CoNb$_3$S$_6$}},\ }\href
{https://doi.org/10.1038/s41467-018-05756-7} {\bibfield  {journal} {\bibinfo
		{journal} {Nat. Commun.}\ }\textbf {\bibinfo {volume} {9}},\ \bibinfo {pages}
	{3280} (\bibinfo {year} {2018})}\BibitemShut {NoStop}%
\bibitem [{\citenamefont {Ishiwata}\ \emph {et~al.}(2011)\citenamefont
	{Ishiwata}, \citenamefont {Tokunaga}, \citenamefont {Kaneko}, \citenamefont
	{Okuyama}, \citenamefont {Tokunaga}, \citenamefont {Wakimoto}, \citenamefont
	{Kakurai}, \citenamefont {Arima}, \citenamefont {Taguchi},\ and\
	\citenamefont {Tokura}}]{Ishiwata2011}%
\BibitemOpen
\bibfield  {author} {\bibinfo {author} {\bibfnamefont {S.}~\bibnamefont
		{Ishiwata}}, \bibinfo {author} {\bibfnamefont {M.}~\bibnamefont {Tokunaga}},
	\bibinfo {author} {\bibfnamefont {Y.}~\bibnamefont {Kaneko}}, \bibinfo
	{author} {\bibfnamefont {D.}~\bibnamefont {Okuyama}}, \bibinfo {author}
	{\bibfnamefont {Y.}~\bibnamefont {Tokunaga}}, \bibinfo {author}
	{\bibfnamefont {S.}~\bibnamefont {Wakimoto}}, \bibinfo {author}
	{\bibfnamefont {K.}~\bibnamefont {Kakurai}}, \bibinfo {author} {\bibfnamefont
		{T.}~\bibnamefont {Arima}}, \bibinfo {author} {\bibfnamefont
		{Y.}~\bibnamefont {Taguchi}},\ and\ \bibinfo {author} {\bibfnamefont
		{Y.}~\bibnamefont {Tokura}},\ }\bibfield  {title} {\bibinfo {title}
	{{Versatile Helimagnetic Phases under Magnetic Fields in Cubic Perovskite
			SrFeO$_3$}},\ }\href {https://doi.org/10.1103/PhysRevB.84.054427} {\bibfield
	{journal} {\bibinfo  {journal} {Phys. Rev. B}\ }\textbf {\bibinfo {volume}
		{84}},\ \bibinfo {pages} {054427} (\bibinfo {year} {2011})}\BibitemShut
{NoStop}%
\bibitem [{\citenamefont {Iguchi}\ \emph {et~al.}(2007)\citenamefont {Iguchi},
	\citenamefont {Hanasaki},\ and\ \citenamefont {Tokura}}]{Iguchi2007}%
\BibitemOpen
\bibfield  {author} {\bibinfo {author} {\bibfnamefont {S.}~\bibnamefont
		{Iguchi}}, \bibinfo {author} {\bibfnamefont {N.}~\bibnamefont {Hanasaki}},\
	and\ \bibinfo {author} {\bibfnamefont {Y.}~\bibnamefont {Tokura}},\
}\bibfield  {title} {\bibinfo {title} {{Scaling of Anomalous Hall Resistivity
			in Nd$_2$(Mo$_{1-x}$Nb$_x$)$_2$O$_7$ with Spin Chirality}},\ }\href
{https://doi.org/10.1103/PhysRevLett.99.077202} {\bibfield  {journal}
	{\bibinfo  {journal} {Phys. Rev. Lett.}\ }\textbf {\bibinfo {volume} {99}},\
	\bibinfo {pages} {077202} (\bibinfo {year} {2007})}\BibitemShut {NoStop}%
\bibitem [{\citenamefont {Kipp}\ \emph {et~al.}(2021)\citenamefont {Kipp},
	\citenamefont {Lux},\ and\ \citenamefont {Mokrousov}}]{Kipp2021b}%
\BibitemOpen
\bibfield  {author} {\bibinfo {author} {\bibfnamefont {J.}~\bibnamefont
		{Kipp}}, \bibinfo {author} {\bibfnamefont {F.~R.}\ \bibnamefont {Lux}},\ and\
	\bibinfo {author} {\bibfnamefont {Y.}~\bibnamefont {Mokrousov}},\ }\bibfield
{title} {\bibinfo {title} {{Chiral response of spin-spiral states as the
			origin of chiral transport fingerprints of spin textures}},\ }\href
{https://doi.org/10.1103/PhysRevResearch.3.043155} {\bibfield  {journal}
	{\bibinfo  {journal} {Phys. Rev. Res.}\ }\textbf {\bibinfo {volume} {3}},\
	\bibinfo {pages} {043155} (\bibinfo {year} {2021})}\BibitemShut {NoStop}%
\bibitem [{\citenamefont {{Lima Fernandes}}\ \emph {et~al.}(2022)\citenamefont
	{{Lima Fernandes}}, \citenamefont {Bl{\"{u}}gel},\ and\ \citenamefont
	{Lounis}}]{LimaF2022}%
\BibitemOpen
\bibfield  {author} {\bibinfo {author} {\bibfnamefont {I.}~\bibnamefont
		{{Lima Fernandes}}}, \bibinfo {author} {\bibfnamefont {S.}~\bibnamefont
		{Bl{\"{u}}gel}},\ and\ \bibinfo {author} {\bibfnamefont {S.}~\bibnamefont
		{Lounis}},\ }\bibfield  {title} {\bibinfo {title} {{Spin-orbit enabled
			all-electrical readout of chiral spin-textures}},\ }\href
{https://doi.org/10.1038/s41467-022-29237-0} {\bibfield  {journal} {\bibinfo
		{journal} {Nat. Commun.}\ }\textbf {\bibinfo {volume} {13}},\ \bibinfo
	{pages} {1576} (\bibinfo {year} {2022})}\BibitemShut {NoStop}%
\bibitem [{\citenamefont {{Y. Wang et al.}}(2022)}]{Wang_2022}%
\BibitemOpen
\bibfield  {author} {\bibinfo {author} {\bibnamefont {{Y. Wang et al.}}},\
}\bibfield  {title} {\bibinfo {title} {{Giant and Reversible Electronic
			Structure Evolution in a Magnetic Topological Material EuCd$_2$As$_2$}},\
}\href {https://doi.org/10.1103/PhysRevB.106.085134} {\bibfield  {journal}
	{\bibinfo  {journal} {Phys. Rev. B}\ }\textbf {\bibinfo {volume} {106}},\
	\bibinfo {pages} {085134} (\bibinfo {year} {2022})}\BibitemShut {NoStop}%
\bibitem [{\citenamefont {Berry}(1984)}]{Berry1984}%
\BibitemOpen
\bibfield  {author} {\bibinfo {author} {\bibfnamefont {M.~V.}\ \bibnamefont
		{Berry}},\ }\bibfield  {title} {\bibinfo {title} {{Quantal Phase Factors
			Accompanying Adiabatic Changes}},\ }\href
{https://doi.org/10.1098/rspa.1984.0023} {\bibfield  {journal} {\bibinfo
		{journal} {Proc. R. Soc. London A}\ }\textbf {\bibinfo {volume} {392}},\
	\bibinfo {pages} {45} (\bibinfo {year} {1984})}\BibitemShut {NoStop}%
\bibitem [{\citenamefont {Xiao}\ \emph {et~al.}(2010)\citenamefont {Xiao},
	\citenamefont {Chang},\ and\ \citenamefont {Niu}}]{Xiao2010}%
\BibitemOpen
\bibfield  {author} {\bibinfo {author} {\bibfnamefont {D.}~\bibnamefont
		{Xiao}}, \bibinfo {author} {\bibfnamefont {M.-C.}\ \bibnamefont {Chang}},\
	and\ \bibinfo {author} {\bibfnamefont {Q.}~\bibnamefont {Niu}},\ }\bibfield
{title} {\bibinfo {title} {{Berry Phase Effects on Electronic Properties}},\
}\href {https://doi.org/10.1103/RevModPhys.82.1959} {\bibfield  {journal}
	{\bibinfo  {journal} {Rev. Mod. Phys.}\ }\textbf {\bibinfo {volume} {82}},\
	\bibinfo {pages} {1959} (\bibinfo {year} {2010})}\BibitemShut {NoStop}%
\bibitem [{\citenamefont {Hu}\ and\ \citenamefont {Rosenbaum}(2008)}]{Hu2008}%
\BibitemOpen
\bibfield  {author} {\bibinfo {author} {\bibfnamefont {J.}~\bibnamefont
		{Hu}}\ and\ \bibinfo {author} {\bibfnamefont {T.~F.}\ \bibnamefont
		{Rosenbaum}},\ }\bibfield  {title} {\bibinfo {title} {{Classical and Quantum
			Routes to Linear Magnetoresistance}},\ }\href
{https://doi.org/10.1038/nmat2259} {\bibfield  {journal} {\bibinfo  {journal}
		{Nat. Mater.}\ }\textbf {\bibinfo {volume} {7}},\ \bibinfo {pages} {697}
	(\bibinfo {year} {2008})}\BibitemShut {NoStop}%
\bibitem [{\citenamefont {Parish}\ and\ \citenamefont
	{Littlewood}(2003)}]{Parish2003}%
\BibitemOpen
\bibfield  {author} {\bibinfo {author} {\bibfnamefont {M.~M.}\ \bibnamefont
		{Parish}}\ and\ \bibinfo {author} {\bibfnamefont {P.~B.}\ \bibnamefont
		{Littlewood}},\ }\bibfield  {title} {\bibinfo {title} {{Non-Saturating
			Magnetoresistance in Heavily Disordered Semiconductors}},\ }\href
{https://doi.org/10.1038/nature02073} {\bibfield  {journal} {\bibinfo
		{journal} {Nature}\ }\textbf {\bibinfo {volume} {426}},\ \bibinfo {pages}
	{162} (\bibinfo {year} {2003})}\BibitemShut {NoStop}%
\bibitem [{\citenamefont {Ali}\ \emph {et~al.}(2016)\citenamefont {Ali},
	\citenamefont {Schoop}, \citenamefont {Garg}, \citenamefont {Lippmann},
	\citenamefont {Lara}, \citenamefont {Lotsch},\ and\ \citenamefont
	{Parkin}}]{Ali2016}%
\BibitemOpen
\bibfield  {author} {\bibinfo {author} {\bibfnamefont {M.~N.}\ \bibnamefont
		{Ali}}, \bibinfo {author} {\bibfnamefont {L.~M.}\ \bibnamefont {Schoop}},
	\bibinfo {author} {\bibfnamefont {C.}~\bibnamefont {Garg}}, \bibinfo {author}
	{\bibfnamefont {J.~M.}\ \bibnamefont {Lippmann}}, \bibinfo {author}
	{\bibfnamefont {E.}~\bibnamefont {Lara}}, \bibinfo {author} {\bibfnamefont
		{B.}~\bibnamefont {Lotsch}},\ and\ \bibinfo {author} {\bibfnamefont
		{S.~S.~P.}\ \bibnamefont {Parkin}},\ }\bibfield  {title} {\bibinfo {title}
	{{Butterfly Magnetoresistance, Quasi-2D Dirac Fermi Surface and Topological
			Phase Transition in ZrSiS}},\ }\href {https://doi.org/10.1126/sciadv.1601742}
{\bibfield  {journal} {\bibinfo  {journal} {Sci. Adv.}\ }\textbf {\bibinfo
		{volume} {2}},\ \bibinfo {pages} {e1601742} (\bibinfo {year}
	{2016})}\BibitemShut {NoStop}%
\bibitem [{\citenamefont {Voerman}\ \emph {et~al.}(2019)\citenamefont
	{Voerman}, \citenamefont {Mulder}, \citenamefont {de~Boer}, \citenamefont
	{Huang}, \citenamefont {Schoop}, \citenamefont {Li},\ and\ \citenamefont
	{Brinkman}}]{Voerman2019}%
\BibitemOpen
\bibfield  {author} {\bibinfo {author} {\bibfnamefont {J.~A.}\ \bibnamefont
		{Voerman}}, \bibinfo {author} {\bibfnamefont {L.}~\bibnamefont {Mulder}},
	\bibinfo {author} {\bibfnamefont {J.~C.}\ \bibnamefont {de~Boer}}, \bibinfo
	{author} {\bibfnamefont {Y.}~\bibnamefont {Huang}}, \bibinfo {author}
	{\bibfnamefont {L.~M.}\ \bibnamefont {Schoop}}, \bibinfo {author}
	{\bibfnamefont {C.}~\bibnamefont {Li}},\ and\ \bibinfo {author}
	{\bibfnamefont {A.}~\bibnamefont {Brinkman}},\ }\bibfield  {title} {\bibinfo
	{title} {{Origin of the Butterfly Magnetoresistance in ZrSiS}},\ }\href
{https://doi.org/10.1103/PhysRevMaterials.3.084203} {\bibfield  {journal}
	{\bibinfo  {journal} {Phys. Rev. Mater.}\ }\textbf {\bibinfo {volume} {3}},\
	\bibinfo {pages} {084203} (\bibinfo {year} {2019})}\BibitemShut {NoStop}%
\bibitem [{\citenamefont {{M. Novak et al.}}(2019)}]{Novak2019}%
\BibitemOpen
\bibfield  {author} {\bibinfo {author} {\bibnamefont {{M. Novak et al.}}},\
}\bibfield  {title} {\bibinfo {title} {{Highly Anisotropic Interlayer
			Magnetoresitance in ZrSiS Nodal-Line Dirac Semimetal}},\ }\href
{https://doi.org/10.1103/PhysRevB.100.085137} {\bibfield  {journal} {\bibinfo
		{journal} {Phys. Rev. B}\ }\textbf {\bibinfo {volume} {100}},\ \bibinfo
	{pages} {085137} (\bibinfo {year} {2019})}\BibitemShut {NoStop}%
\bibitem [{\citenamefont {Pavlosiuk}\ \emph {et~al.}(2020)\citenamefont
	{Pavlosiuk}, \citenamefont {Fałat}, \citenamefont {Kaczorowski},\ and\
	\citenamefont {Wiśniewski}}]{Pavlosiuk2020}%
\BibitemOpen
\bibfield  {author} {\bibinfo {author} {\bibfnamefont {O.}~\bibnamefont
		{Pavlosiuk}}, \bibinfo {author} {\bibfnamefont {P.}~\bibnamefont {Fałat}},
	\bibinfo {author} {\bibfnamefont {D.}~\bibnamefont {Kaczorowski}},\ and\
	\bibinfo {author} {\bibfnamefont {P.}~\bibnamefont {Wiśniewski}},\
}\bibfield  {title} {\bibinfo {title} {{Anomalous Hall Effect and Negative
			Longitudinal Magnetoresistance in Half-Heusler Topological Semimetal
			Candidates TbPtBi and HoPtBi}},\ }\href {https://doi.org/10.1063/5.0026956}
{\bibfield  {journal} {\bibinfo  {journal} {APL Mater.}\ }\textbf {\bibinfo
		{volume} {8}},\ \bibinfo {pages} {111107} (\bibinfo {year}
	{2020})}\BibitemShut {NoStop}%
\bibitem [{\citenamefont {Huynh}\ \emph {et~al.}(2019)\citenamefont {Huynh},
	\citenamefont {Ogasawara}, \citenamefont {Kitahara}, \citenamefont {Tanabe},
	\citenamefont {Matsushita}, \citenamefont {Tahara}, \citenamefont {Kida},
	\citenamefont {Hagiwara}, \citenamefont {Arčon},\ and\ \citenamefont
	{Tanigaki}}]{Huynh2019}%
\BibitemOpen
\bibfield  {author} {\bibinfo {author} {\bibfnamefont {K.-K.}\ \bibnamefont
		{Huynh}}, \bibinfo {author} {\bibfnamefont {T.}~\bibnamefont {Ogasawara}},
	\bibinfo {author} {\bibfnamefont {K.}~\bibnamefont {Kitahara}}, \bibinfo
	{author} {\bibfnamefont {Y.}~\bibnamefont {Tanabe}}, \bibinfo {author}
	{\bibfnamefont {S.~Y.}\ \bibnamefont {Matsushita}}, \bibinfo {author}
	{\bibfnamefont {T.}~\bibnamefont {Tahara}}, \bibinfo {author} {\bibfnamefont
		{T.}~\bibnamefont {Kida}}, \bibinfo {author} {\bibfnamefont {M.}~\bibnamefont
		{Hagiwara}}, \bibinfo {author} {\bibfnamefont {D.}~\bibnamefont {Arčon}},\
	and\ \bibinfo {author} {\bibfnamefont {K.}~\bibnamefont {Tanigaki}},\
}\bibfield  {title} {\bibinfo {title} {{Negative and Positive
			Magnetoresistance in the Itinerant Antiferromagnet BaMn$_2$\textit{Pn}$_2$
			(\textit{Pn} = P, As, Sb, and Bi)}},\ }\href
{https://doi.org/10.1103/PhysRevB.99.195111} {\bibfield  {journal} {\bibinfo
		{journal} {Phys. Rev. B}\ }\textbf {\bibinfo {volume} {99}},\ \bibinfo
	{pages} {195111} (\bibinfo {year} {2019})}\BibitemShut {NoStop}%	
\end{thebibliography}
%%\end{document}

\clearpage
%\newpage
\onecolumngrid
\begin{center}
	\textbf{Large unconventional anomalous Hall effect arising from spin chirality within~domain~walls~of~an~antiferromagnet EuZn$_2$Sb$_2$}\\[.2cm]
	K. Singh, O. Pavlosiuk, S. Dan, D. Kaczorowski, P. Wiśniewski \\[.2cm]	
	{\itshape Institute of Low Temperature and Structure Research,\\ Polish Academy of Sciences, Wrocław, Poland}\\[.5cm]
	\textbf{\Large Supplemental Material }\\[.3cm]
	%%%%%%
	%%%%%%
	%(Dated: \today)\\[1cm]
\end{center}\vspace{0.2cm}
\setcounter{equation}{0}
\renewcommand{\theequation}{EqS\arabic{equation}}
\setcounter{figure}{0}
\renewcommand{\thefigure}{S\arabic{figure}}
\setcounter{section}{0}
\renewcommand{\thesection}{S\arabic{section}}
\setcounter{table}{0}
\renewcommand{\thetable}{S\arabic{table}}
\renewcommand{\thepage}{S-\arabic{page}}
\setcounter{page}{1}
%%%%%%%%%%%%%%%%%
%%%%%%%%%%%%%%%%%%
\begin{center}
	\textbf{Sample characterization}
\end{center}
Energy dispersive X-ray spectra (EDS) on a few pieces of the synthesized EuZn$_2$Sb$_2$ single crystals show that stoichiometry is in a good agreement with the nominal composition (see Fig.~S1a). The back-scattering Laue diffraction (see Fig.~S1b) confirmed that EuZn$_2$Sb$_2$ crystallizes in the trigonal structure with space-group $P\bar{3}m1$, as it has been reported previously \cite{Weber2006}. This crystal structure is sketched in Figure S2a-c.
\begin{figure*}[h]
	\centering
	\includegraphics[width=\textwidth]{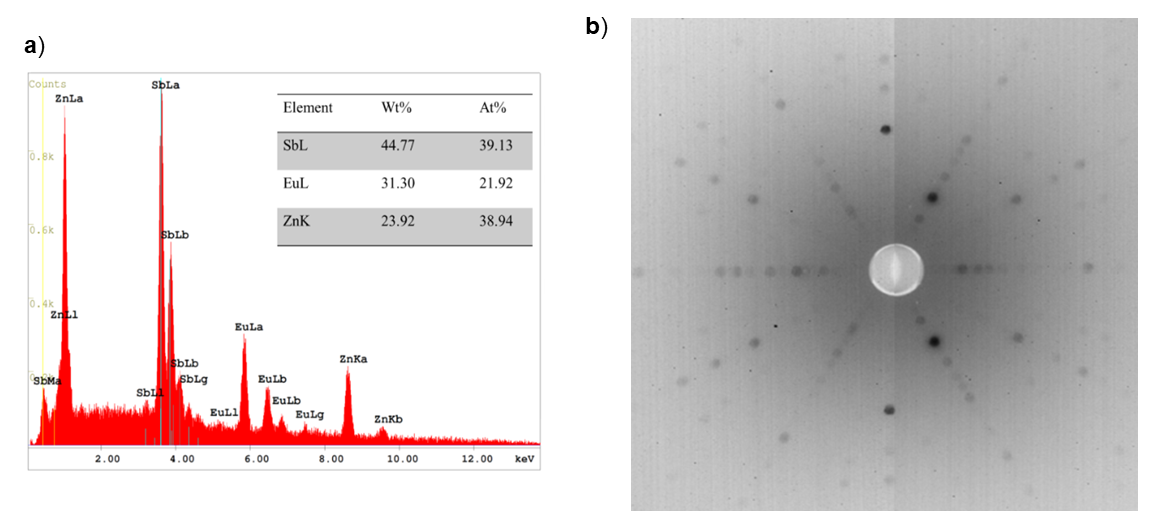}
	\caption{({\bf a}) EDS spectrum along with resulting weight and atomic fractions of EuZn$_2$Sb$_2$. ({\bf b}) Backscattering Laue diffraction pattern taken with incident X-ray beam along [001] crystallographic direction.}
	\label{FigS1}
\end{figure*}
%%%%%%
\begin{figure*}
	\centering
	\includegraphics[width=0.8\textwidth]{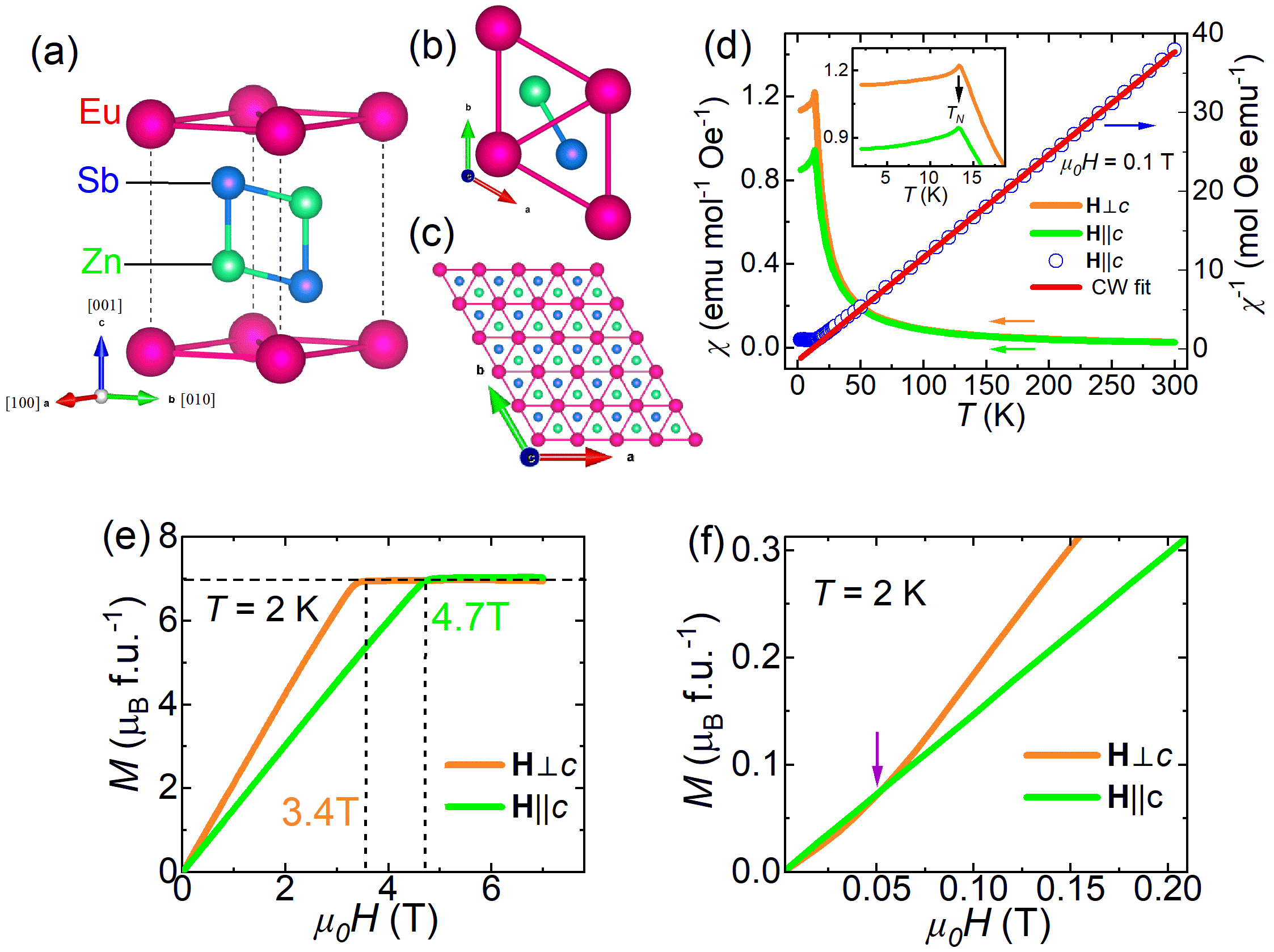}
	\caption{({\bf a}) Crystal structure of EuZn$_2$Sb$_2$; ({\bf b}) and ({\bf c}) Projections of the unit cell and of crystal structure on the hexagonal \textit{ab}-plane, respectively. 
		({\bf d}) Left axis: temperature dependence of the zero-field-cooled magnetic susceptibility in a magnetic field of 0.1\,T, applied in the $ab$-plane and along the ${c}$-axis (orange and green curve, respectively). 
		Inset: low temperature region of {\bf d}, black arrow marks $T_{\rm N}$. Right axis: temperature dependent inverse magnetic susceptibility. The red line is the Curie-Weiss law fit. 
		({\bf e}) Field dependence of magnetization, at $T$=2\,K, with the magnetic field parallel to the $ab$-plane (orange curve) and ${c}$-axis (green curve). ({\bf f}) Low field region of $M(H)$ at 2K, purple arrow indicates the spin-flop transition.}
	\label{FigS2}
\end{figure*}
%%%%%%
\begin{center}
	\textbf{Magnetic properties}
\end{center}
Figure S2d (left axis) shows the temperature dependent magnetic susceptibility, $\chi$, obtained for magnetic field ($H$) of 0.1\,T, which was applied perpendicular or parallel to the \textit{c}-axis. It clearly shows the antiferromagnetic phase transition at $T_{\rm N}$ = 13.3\,K, exactly the same as reported previously \cite{Weber2006}. In the paramagnetic region, the data obtained for $\textbf{H}\parallel{c}$ were fitted with the Curie-Weiss law (red solid line in Fig.~S2d): $\chi$ = $C/(T-\theta)+\chi_0$; where $\chi_0$ is the temperature independent magnetic susceptibility, $C$ is the Curie constant and $\theta$ is the paramagnetic Curie temperature. The effective magnetic moment ($\mu_{eff}$)  was extracted from the Curie constant and equals to 7.95 $\mu_B$, a value almost identical to the theoretical value of 7.94 $\mu_B$ for Eu$^{2+}$. The obtained value of $\theta$ is 11.7\,K, which indicates the dominance of ferromagnetic interactions in the studied material, and is slightly larger than 8.8\,K reported earlier \cite{Weber2006}. We also noticed that the ratio $\chi_\perp$/$\chi_\parallel$ (where $\chi_\perp$ and $\chi_\parallel$ correspond to magnetic susceptibility measured in $\textbf{H}\perp{c}$ and $\textbf{H}\parallel{c}$) increases rapidly with decreasing temperature below 15\,K and reaches a value of 1.33 at $T$ = 2\,K for $\mu_0H$ = 0.1\,T, indicating spin alignment in the ${ab}$-plane, in perfect agreement with previous report \cite{Weber2006}.
Magnetization as a function of magnetic field was measured at $T$ = 2\,K, for $\textbf{H}\parallel{c}$ and $\textbf{H}\perp{c}$ (Fig.~S2e). With increasing magnetic field, $M$ increases and saturates at a 7.2 $\mu_B$ for both field directions. In addition, we observed a difference between the $M(H)$ measured for $\textbf{H}\parallel{c}$  and $\textbf{H}\perp{c}$, which attain saturation for $\mu_0H_{\rm sat}$ $>$ 4.7\,T and 3.5\,T, respectively. In the case of $\textbf{H}\perp{c}$, a low-field (in ca. 50 mT) spin-flop transition is observed in the magnetization (see Fig.~S2f). For $\textbf{H}\parallel{c}$, spin-flop transition was not observed. All these observations are in agreement with those of Ref.~\cite{Weber2006}.
%%%%%%
\newpage
\begin{center}
	\textbf{Heat capacity}
\end{center}
\begin{figure*}
	\centering
	\includegraphics[width=0.6\textwidth]{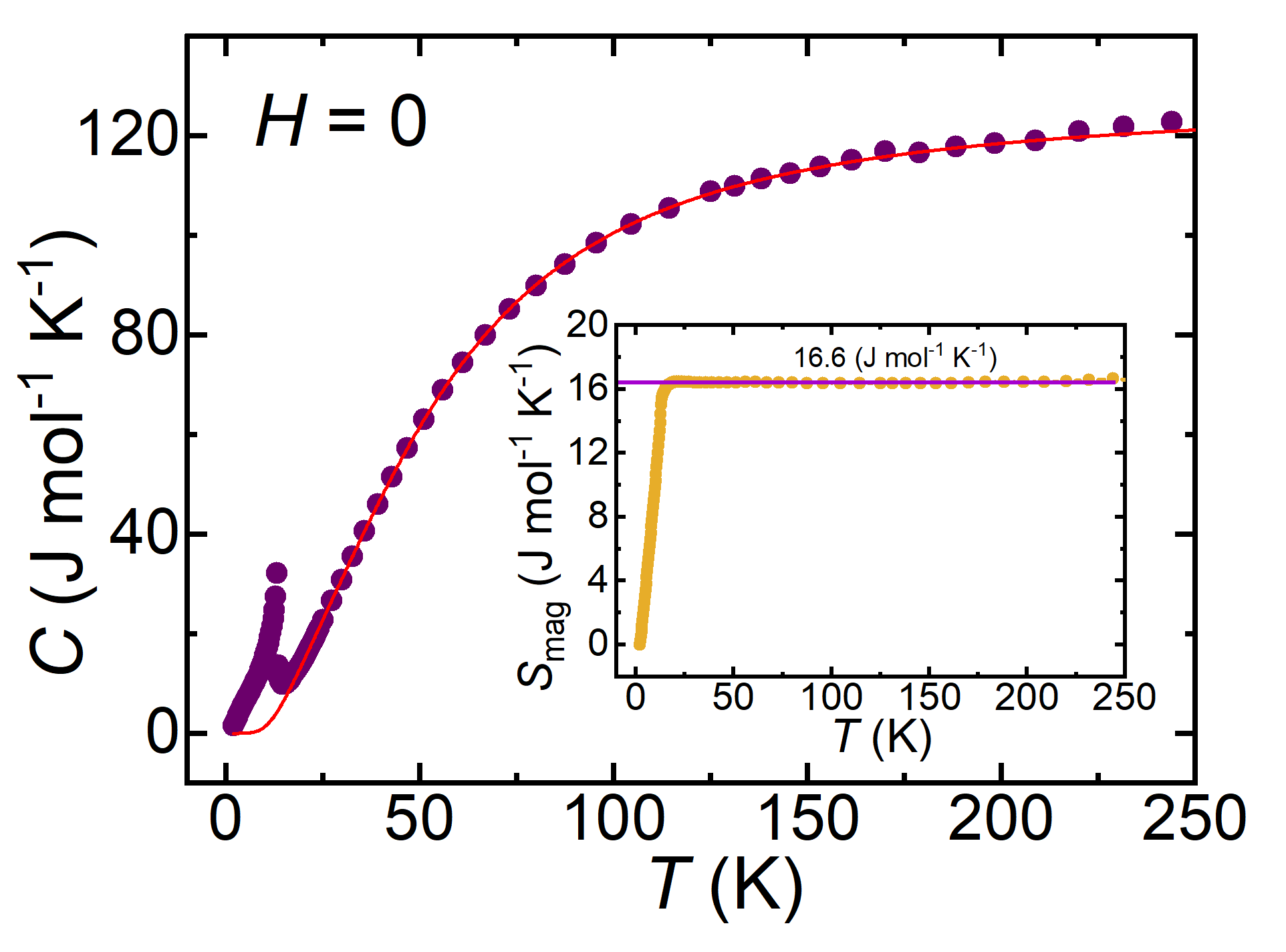}
	\caption{Temperature dependent heat capacity $C_p$ in zero magnetic field. Red curve shows the fit with eq.~\ref{eq_C}. Inset: Temperature dependent magnetic entropy $S_{mag}$.}
	\label{FigS3}
\end{figure*}
%%%%%%
Antiferromagnetic order in EuZn$_2$Sb$_2$ was confirmed by the specific heat ($C_p$) measurement, which is in good agreement with that described in Ref.~\cite{May2012}. The peak of the $\lambda$-shaped anomaly in $C_p(T)$ is located at $T=13.1\,{\rm K}$ (see Fig.~\ref{FigS3}), which is very close to $T_{\rm N}$ obtained from the magnetization measurements. Collected data were fitted using the equation:
\begin{equation}
	C_p(T) = \gamma T + (1-m) C_D(T) + m C_E(T),
	\label{eq_C} 	
\end{equation}  
where $m$ is weight of the Einstein term and $\gamma$ is the Sommerfeld coefficient. $C_D$ and $C_E$ are the Debye model and Einstein model contributions to the lattice heat capacity, described by:
$$
C_D(T) = 9n\mathcal{R}\left(\frac{T}{\theta_{\rm D}}\right)^3 \int_{0}^{{\,\theta_{\rm D}}/{T}}\frac{x^4e^x}{{{(e}^x-1)}^2}dx\quad{\rm and}\quad     
C_E(T) = 3n\mathcal{R} \left(\frac{\theta_{\rm E}}{T}\right)^2 \frac{e^{{\theta_{\rm E}}/{T}}}{\left(e^{{\theta_{\rm E}}/{T}}-1\right)^2}
\label{eqn:cp},  $$        
respectively, where $n$ is the number of atoms per formula unit, $\mathcal{R}$ is universal gas constant, $\theta_D$ and $\theta_E$ are Debye temperature and Einstein temperature, respectively. 
The fitting parameters $m$, $\gamma$, $\theta_D$ and $\theta_E$ equal to $0.260(1)$, $0.0043(3)\,{\rm Jmol}^{-1}{\rm K}^{-2}$, $247.68(7)\,$K, and $77.31(2)\,$K, respectively. The magnetic entropy ($S_{mag}$) was calculated as the integral: %$\textit{S}_{mag} = \int_{2\,\rm K}^{300\,\rm K}{(\frac{\textit{C}}{T}-(\frac{\textit{C}_{ph}}{T}} - \gamma)) {\rm d}T$, 
$S_{mag} = \int_{2\,\rm K}^{300\,\rm K}{((C_p/T)-(C_{ph}/T}) - \gamma)\,{\rm d}T$, where $C_{ph}$ is total phonon heat capacity due to Debye and Einstein contributions (see inset of Fig.~\ref{FigS3}). 
The calculated magnetic entropy, $S_{mag}$, increases with increasing temperature, reaches a maximum of $16.6\,{\rm Jmol}^{-1}{\rm K}^{-2}$ just above $T_{\rm N}$, and then at higher $T$ is almost temperature-independent. The maximum value is very close to the theoretical $S_{mag}=\mathcal{R}\,$ln(2S+1) = $17.3\,{\rm Jmol}^{-1}{\rm K}^{-2}$ for Eu$^{2+}$ (S = 7/2). 

\begin{figure*}[b]
	\centering
	\includegraphics[width=0.6\textwidth]{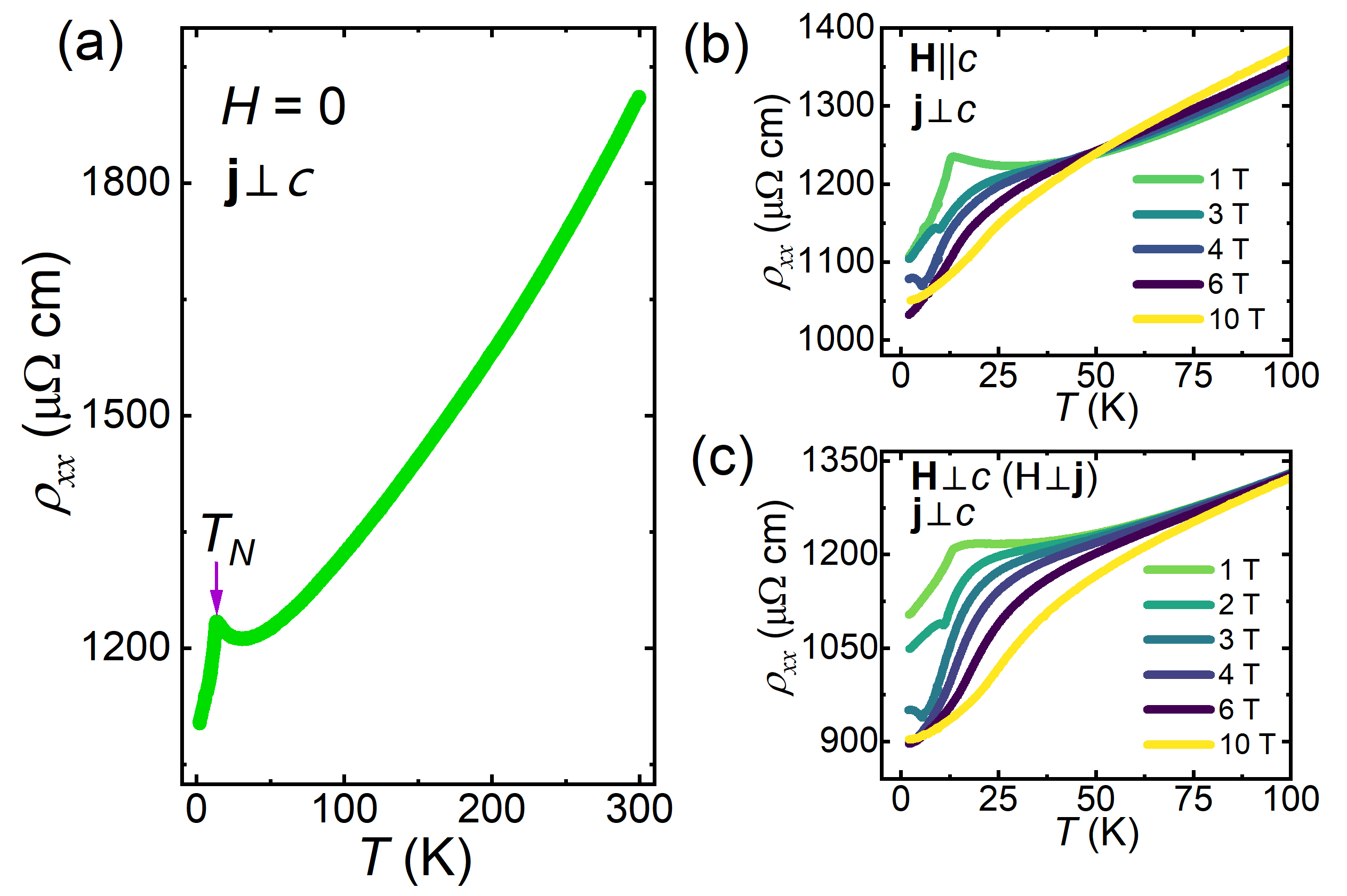}
	\caption{({\bf a}) Temperature dependence of electrical resistivity $\rho_{xx}$ at zero magnetic field. The purple arrow marks the Néel temperature. ({\bf b}) and ({\bf c}): Temperature dependent transverse resistivity in the field range of 1--10\,T, for $\textbf{H}\parallel{c}$ and $\textbf{H}\perp{c}$, respectively.}
	\label{FigS4}
\end{figure*}

%%%%%%
\newpage
\begin{center}
	\textbf{Electrical resistivity}
\end{center}
Figure~\ref{FigS4}a shows the temperature-dependent electrical resistivity ($\rho_{xx}$), measured at zero magnetic field and with electrical current ($j$) in the $ab$ plane. Our data are nearly identical to those reported previously for single-crystalline EuZn$_2$Sb$_2$ \cite{May2012}. The residual resistivity ratio value (RRR = $\rho_{xx}(300\,{\rm K})/\rho_{xx}(2\,{\rm K})$) is 1.7. In the temperature range from 2 to 100\,K, $\rho_{xx}(T)$ was also measured in different magnetic fields, applied parallel or perpendicular to ${c}$-axis (see Figs.~\ref{FigS4}b and \ref{FigS4}c). For $\textbf{H}\perp{c}$ and for $\textbf{H}\parallel{c}$, $T_{\rm N}$ decreases with increasing magnetic field, however for $\textbf{H}\parallel{c}$ the decrease is weaker, reflecting the magnetic anisotropy in the studied compound.
%%%%%%

\newpage
\begin{center}
	\textbf{Temperature variation of normal and anomalous Hall resistivity}
\end{center}
\begin{figure*}[h]
	\centering
	\includegraphics[width=0.6\textwidth]{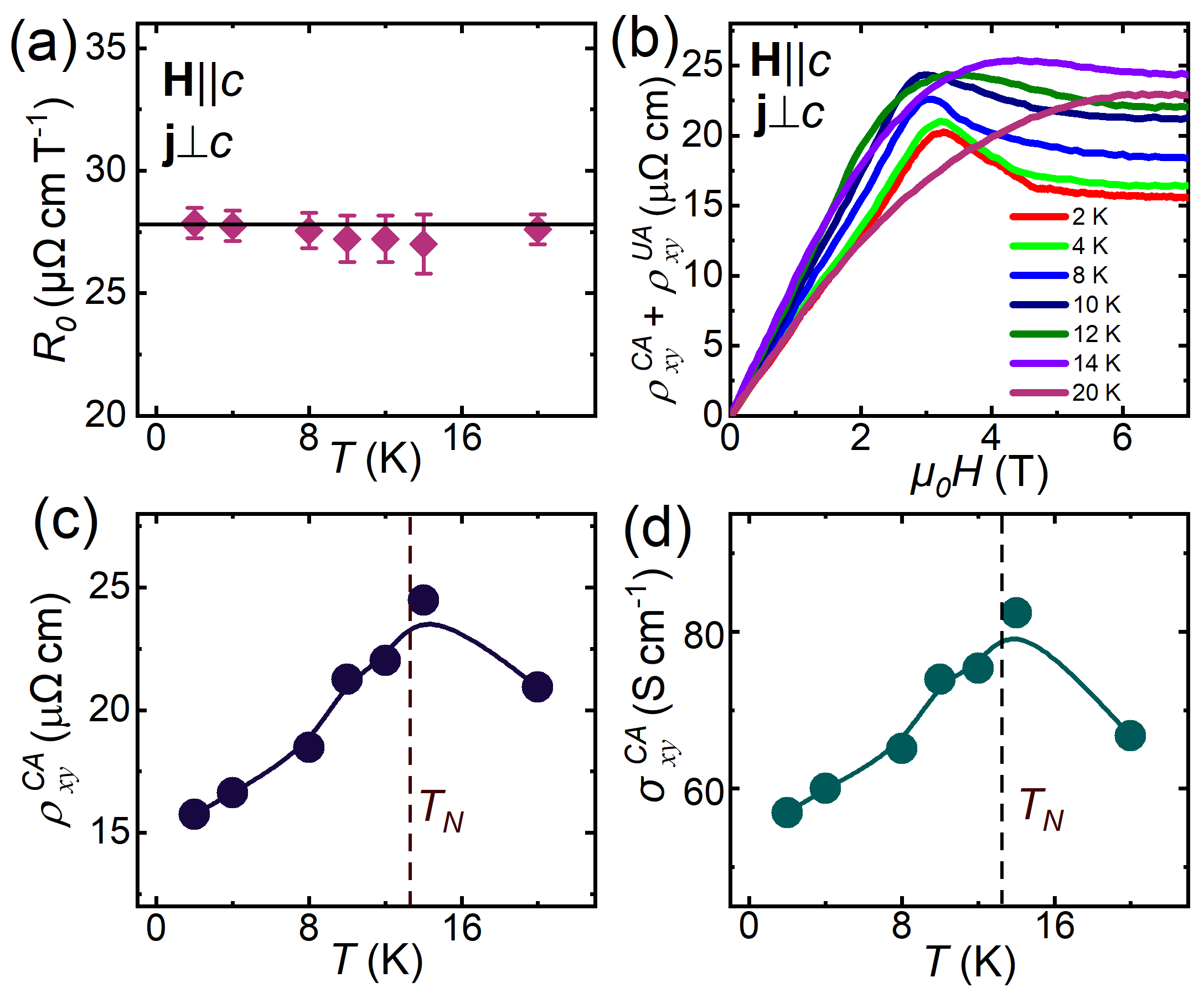}
	\caption{({\bf a}) Temperature dependence of normal Hall coefficient $R_0$. ({\bf b}) Magnetic field dependence of $\rho^{CA}_{xy}$+ $\rho^{\rm UA}_{xy}$ at different temperatures. ({\bf c}) Temperature dependence of conventional anomalous Hall effect, $\rho^{CA}_{xy}$, in field of 7\,T. ({\bf d}) Temperature dependence of conventional anomalous Hall conductivity, $\sigma^{CA}_{xy}(= {\rho^{CA}_{xy}}/{(\rho^2_{xx}+\rho^2_{xy})})$, in field of 7 T}
	\label{FigS5}
\end{figure*}

\end{document}